\documentclass[a4paper,12pt]{article}
\usepackage[DIV12]{typearea}

\usepackage[font=small,labelfont=bf]{caption}
\usepackage{graphicx}
\usepackage{amsmath,amssymb,amsfonts}
\usepackage{verbatim}
\usepackage{bm}
\usepackage{color}
\usepackage{ulem}
\usepackage{mathtools}
\usepackage{cite}
\usepackage{colortbl}
\definecolor{light-gray}{gray}{0.85}
\usepackage[pdftex,colorlinks,pdfpagelabels]{hyperref}
\hypersetup{
   bookmarksnumbered,
   citecolor={blue},
   linkcolor={blue},
   urlcolor={blue},
   filecolor={blue}
} 
\newcommand{\eVdist}{\kern-0.06em}

\newcommand{\gev}{\:\text{Ge\eVdist V}}
\newcommand{\mb}{\:\text{mb}}

\newcommand{\pl}{p_{\text{\tiny{L}}}}
\newcommand{\pt}{p_{\text{\tiny{T}}}}
\newcommand{\mt}{m_{\text{\tiny{T}}}}
\newcommand{\iso}{\ensuremath{\Delta_{\text{IS}}}}
\newcommand{\hyp}{\ensuremath{\Delta_{\Lambda}}}

\begin{document}
\thispagestyle{empty}
\begin{flushright}
NORDITA-2017-6
\end{flushright}

\vspace*{1.0cm}
\begin{center}

{\LARGE\bf Cosmic Ray Antiprotons at High Energies}\\[12mm]

{\large Martin Wolfgang Winkler} 
\\[6mm]
{\it Nordita, KTH Royal Institute of Technology and Stockholm University\\
Roslagstullsbacken 23, 10 691 Stockholm, Sweden
}
\vspace*{12mm}
\begin{abstract}
Cosmic ray antiprotons provide a powerful tool to probe dark matter annihilations in our galaxy. The sensitivity of this important channel is, however, diluted by sizable uncertainties in the secondary antiproton background. In this work, we improve the calculation of secondary antiproton production with a particular focus on the high energy regime. We employ the most recent collider data and identify a substantial increase of antiproton cross sections with energy. This increase is driven by the violation of Feynman scaling as well as by an enhanced strange hyperon production. The updated antiproton production cross sections are made publicly available for independent use in cosmic ray studies. In addition, we provide the correlation matrix of cross section uncertainties for the AMS-02 experiment. At high energies, the new cross sections improve the compatibility of the AMS-02 data with a pure secondary origin of antiprotons in cosmic rays.
\end{abstract}
\end{center}
\clearpage

\section{Introduction}
The last decades have brought tremendous improvements in the measurement of the cosmic ray antiproton flux. With the latest AMS-02 data set, uncertainties decreased to the level of $5\%$ over a wide range of energies~\cite{Aguilar:2016kjl}. This opens new opportunities to pin down the propagation properties of charged cosmic rays. In addition, it implies a unique sensitivity to any extra source of antiprotons, in particular to dark matter annihilations. Indeed, thermal dark matter candidates with masses up to several hundred GeV are potentially accessible to AMS-02~\cite{Giesen:2015ufa,Cuoco:2016eej}. This makes antiprotons one of the most powerful channels for indirect dark matter detection.

Unfortunately, the calculation of the secondary antiproton background -- the latter stems from scattering of primary cosmic rays on the interstellar matter -- has not kept up with the precision of the measurement. At this stage, any conclusion drawn from the observed flux is limited by the background uncertainty rather than the experimental error. Improvement of the background modeling appears even more urgent in the light of the experimental situation: when the preliminary AMS-02 antiproton flux was released in 2015, the spectrum appeared to be surprisingly hard~\cite{tingtalk:2015}. Subsequent studies which employed updated propagation parameters strongly improved the compatibility of the measured flux with the background prediction~\cite{Evoli:2015vaa,Kappl:2015bqa}. Still, the observed slope appeared towards the edge of the uncertainty band. This lead to the question, whether there is an additional systematic effect which hardens the antiproton spectrum. In this work we show that such an effect exists in the antiproton cross sections. 

In a previous study we determined the cross sections within the regime of radial/ Feynman scaling~\cite{Kappl:2014hha}. In this work, we identify sources of scaling violation which become relevant at high energies. This is done by using precision experimental data on proton proton scattering from the STAR and PHENIX experiments at RHIC as well as from ALICE and CMS at LHC. With growing energy we also observe a relative increase of the strange hyperon multiplicity which implies that hyperon-induced antiprotons gain significance. 

A complementary issue raised in~\cite{Fischer:2003xh,diMauro:2014zea,Kappl:2014hha} regards the possible enhancement of antineutron over antiproton production due to isospin effects. We refine our previous study by using experimental data on proton neutron and proton nucleus scattering as well as symmetry arguments. 
This enables us to determine the energy-dependence of the isospin factor and to include it in our revised calculation of cross sections. Our new antiproton cross sections are made publicly available.

In order to make contact with observation, we, furthermore, determine the antiproton flux within the standard two-zone diffusion model. In particular, we quantify the increase of the high energy antiproton flux which follows from our new cross section evaluation. We keep track of all relevant cross section uncertainties and their correlations. This allows us to rigorously test the consistency of the AMS-02 data with pure secondary production of cosmic ray antiprotons.

\section{Antiproton Production Channels}

Cosmic ray antiprotons mainly stem from the scattering of protons and helium on the interstellar matter. Proton proton scattering constitutes the dominant process and shall be the main focus of this work. We follow our previous study~\cite{Kappl:2014hha} and separate the total antiproton production cross section $\sigma$ into several pieces
\begin{equation}
  \sigma = \sigma_{\bar{p}} + \sigma_{\bar{n}}\,,\qquad \sigma_{\bar{p},\bar{n}} = \sigma^0_{\bar{p},\bar{n}} + \sigma^{\Lambda}_{\bar{p},\bar{n}}\,.
\end{equation}
In the first step we distinguished between antiprotons which are produced on short time-scales (index $\bar{p}$) and those from the late decays of antineutrons (index $\bar{n}$). Both pieces are further separated depending on whether the antiprotons (or antineutrons) stem directly from the factorization of the colliding partons (superscript $0$) or from the decay of intermediate strange hyperons (superscript $\Lambda$). We comment that hyperons have a decay length comparable to typical detector scales and therefore require a separate treatment.

It is convenient to introduce the (Lorentz) invariant differential cross section for antiproton production in proton proton scattering
\begin{equation}
f= E \frac{d^3\sigma}{dp^3}= \frac{E}{\pi} \frac{d^2\sigma}{d\pl d\pt^2}\,,
\end{equation}
with $E$, $\pl$ and $\pt$ denoting the antiproton energy, longitudinal and transverse momentum. Another useful quantity is the antiproton multiplicity which is defined as the average number of antiprotons generated per inelastic proton proton scattering event, i.e.
\begin{equation}
  n = \frac{\sigma}{\sigma_\text{in}}\,,
\end{equation} 
where $\sigma_{\text{in}}$ stands for the total inelastic cross section. We will use the same sub- and superscripts as above, e.g. $n_{\bar{p}}^0$ refers to the multiplicity of promptly produced antiprotons. Furthermore, we introduce the radial and Feynman scaling variables which are defined as
\begin{equation}
 x_R = \frac{E^*}{E_{\text{max}}^*}\,, \qquad  x_f = \frac{\pl^*}{\sqrt{s}/2}\,,
\end{equation}
where $E^*$ and $\pl^*$ are the energy and longitudinal momentum in the center-of-mass frame. The maximal energy is determined as $E_{\text{max}}^*=(s-8 m_p^2)/(2\sqrt{s})$ with the proton mass $m_p$. For a range of energies $10\gev \lesssim \sqrt{s} \lesssim 50 \gev$, the invariant cross section $f$ is (approximately) independent of $\sqrt{s}$ if expressed in terms of $\pt$ and a scaling variable (see section~\ref{sec:scaling}). 

We now turn to the determination of the invariant cross section which we rewrite in the form
\begin{equation}\label{eq:contributions}
  f= f_{\bar{p}}^0 \:( 2 + \iso + 2 \,\hyp)\,,
\end{equation}
where we defined the isospin enhancement factor $\iso = f_{\bar{n}}^0 / f_{\bar{p}}^0 -1$, the hyperon factor $\hyp = f_{\bar{p}}^\Lambda / f_{\bar{p}}^0$ and assumed $f_{\bar{p}}^\Lambda=f_{\bar{n}}^\Lambda$.\footnote{In~\cite{Kappl:2014hha} symmetry and isospin arguments were used to estimate that $f_{\bar{n}}^\Lambda$ would deviate by a few per cent from $f_{\bar{n}}^\Lambda$. This small difference can be neglected as hyperon-induced processes only make up a fraction of the total antiproton cross section.} 

The next section is devoted to the determination of $\hyp$ while we will discuss $f_{\bar{p}}^0$ in sections~\ref{sec:scaling} and~\ref{sec:violation}, before turning to $\iso$ in section~\ref{sec:isospin}.

\section{Hyperons}\label{sec:hyperon}

In hadronic scattering processes a sizable fraction of antiprotons is produced by decay of the antihyperons $\bar{\Lambda}$ and $\bar{\Sigma}$. As hyperons have a macroscopic decay length $c\tau\gtrsim \text{cm}$ in the detector it is not obvious that the daughter antiprotons contribute to the cross section measured at an accelerator experiment. Indeed, most present collider experiments apply a feed-down correction to their data, i.e.\ they use precision tracking techniques to reject antiprotons from hyperon decay. The situation is further complicated by the fact that older experimental data from the 1970s and 80s do not contain such a feed-down correction and, hence, a comparison is not straightforward.

\begin{figure}[htp]
\begin{center}   
 \includegraphics[width=15cm]{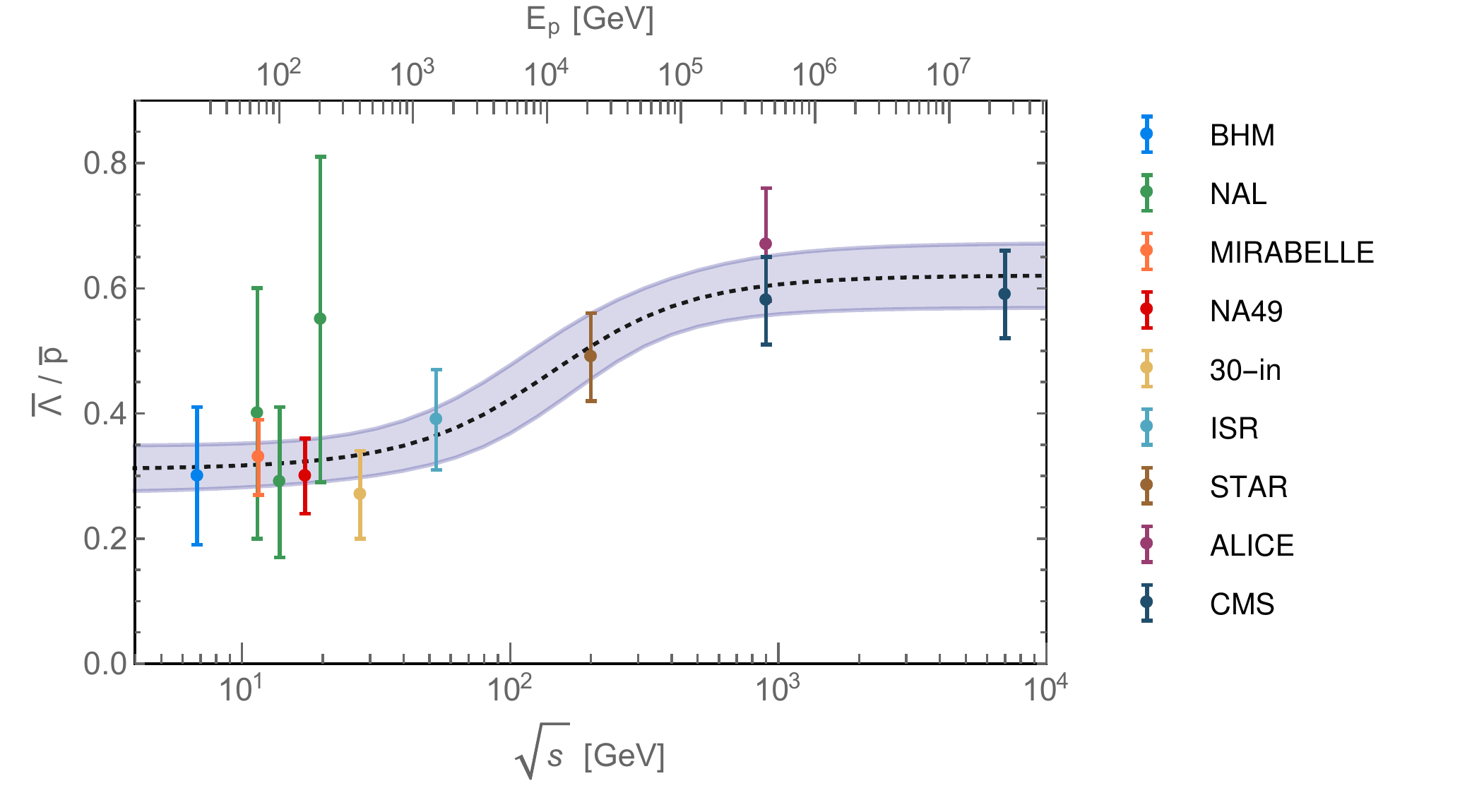}
\end{center}
\caption{$\bar{\Lambda}/\bar{p}$ ratio in proton proton collisions as measured by several experiments (see text). The fit with the parameterization~\eqref{eq:lambdapara} and the corresponding uncertainty band are also shown.}
\label{fig:hyperon}
\end{figure}

In this section we aim at determining explicitly the ratio $\hyp$ of hyperon-induced to promptly produced antiprotons in proton proton scattering. As experimental data indicate that the phase space distributions of the resulting antiprotons matches between the two production modes~\cite{Alt:2005zq,Kappl:2014hha}, we take $\hyp$ to be only a function of the center-of-mass energy $\sqrt{s}$. We can extract it from the parent hyperon production
\begin{equation}
  \hyp = \frac{\bar{\Lambda}}{\bar{p}} \times \text{Br}\left(\bar{\Lambda}\rightarrow \bar{p} + \pi^+\right)
  +  \frac{\bar{\Sigma}^-}{\bar{p}} \times \text{Br}\left(\bar{\Sigma}^-\rightarrow \bar{p} + \pi^0\right)\,,
\end{equation}
where $\bar{\Lambda}/\bar{p}$ and ${\bar{\Sigma}}^-/\bar{p}$ are the hyperon to (promptly produced) antiproton ratios. Due to lack of experimental data on $\bar{\Sigma}$ production, we follow~\cite{Kappl:2014hha} and use symmetry arguments to estimate $\bar{\Sigma}/\bar{\Lambda}= 0.33$. We assume a $25\%$ uncertainty on this ratio. Taking the branching fractions from~\cite{Olive:2016xmw} we arrive at 
\begin{equation}\label{eq:convert}
\hyp = (0.81\pm 0.04)\,(\bar{\Lambda}/\bar{p})\,.
\end{equation} 
The following data sets are used to determine $\bar{\Lambda}/\bar{p}$ as a function of energy:
\begin{itemize}
\item
the Bonn-Hamburg-M\"unchen (BHM) collaboration measured $\sigma_{\bar{\Lambda}}=(0.021\pm 0.007)$ mb at $\sqrt{s}=6.8\gev$~\cite{Blobel:1973jc} which is combined with $\sigma_{\bar{p}}=(0.086\pm 0.014)\mb$~\cite{Amaldi:1975hot} to yield $\bar{\Lambda}/\bar{p} =0.30\pm 0.11$. There was no feed-down correction applied in~\cite{Amaldi:1975hot} and we thus used~\eqref{eq:convert} to arrive at the given ratio.
\item
the cross sections $\sigma_{\bar{\Lambda}}=0.20\pm 0.10,\; 0.23\pm 0.10,\; 0.83\pm 0.39\mb$ were obtained with the NAL hydrogen bubble chamber at $\sqrt{s}=11.4,\; 13.8,\; 19.6\gev$~\cite{Whitmore:1973ri}. The prompt antiproton cross section was not measured at the same energies. However, as these energies reside in the radial scaling window, we can extract it from~\cite{Kappl:2014hha}. This leads to the ratios $\bar{\Lambda}/\bar{p} =0.40\pm 0.20,\; 0.29 \pm 0.12,\; 0.55 \pm 0.26$.
\item
the France-Soviet Union Collaboration measured $\sigma_{\bar{\Lambda}}=0.16\pm 0.03\mb$ at $\sqrt{s}=11.5\gev$ with the MIRABELLE hydrogen bubble chamber~\cite{Ammosov:1975bt}. Again extracting the antiproton cross section from~\cite{Kappl:2014hha}, we obtain $\bar{\Lambda}/\bar{p} =0.33\pm 0.06$.
\item
the NA49 collaboration published the differential multiplicity of $\bar{\Lambda}$ at $\sqrt{s}=17.2\gev$ in~\cite{Alt:2005zq}. Integrating the distribution we obtain $n_{\bar{\Lambda}} = 0.0117$. As there are no error bars given in~\cite{Alt:2005zq} we estimate an uncertainty of $20\%$ from the uncertainty in the feed-down correction stated in~\cite{Anticic:2009wd}. In the same reference one finds $n_{\bar{p}}^0= 0.0386 \pm 0.0025$, such that $\bar{\Lambda}/\bar{p} =0.30\pm 0.06$.
\item
in~\cite{Kichimi:1979te} data from the Fermilab 30-in bubble chamber were analyzed. It was found that $\sigma_{\bar{\Lambda}}=0.63\pm 0.12$ at $\sqrt{s}=27.6\gev$. Again using~\cite{Kappl:2014hha} we translate this into $\bar{\Lambda}/\bar{p}=0.27\pm 0.07$.
\item 
the CERN ISR data shown in~\cite{Fischer:2003xh} indicate that $\int d \pt^2\, f_{\bar{\Lambda}}= 0.156\pm 0.030\mb$ at $\sqrt{s}=53\gev$ and $x_F=0$. We use the measurement of the invariant cross section by the British-Scandinavian collaboration~\cite{Alper:1975jm} to evaluate the same quantity for antiprotons and find $\int d \pt^2\, f_{\bar{p}}= 0.52\pm 0.02\mb$. As no feed-down correction was applied to~\cite{Alper:1975jm} we use~\eqref{eq:convert} to arrive at $\Lambda/\bar{p} =0.39\pm 0.08$. 
\item
the STAR collaboration measured the multiplicities of $\bar{p}$ and $\bar{\Lambda}$ at central rapidity $y$ in $\sqrt{s}=200\gev$ proton proton collisions. The results $dn_{\bar{\Lambda}}/dy|_{y=0} = 0.0398 \pm 0.0038$~\cite{Abelev:2006cs} and $dn_{\bar{p}}/dy|_{y=0} = 0.113\pm 0.010$~\cite{Abelev:2008ab} translate into $\bar{\Lambda}/\bar{p} =0.49\pm 0.07$. As STAR did not apply a feed-down correction we have used~\eqref{eq:convert}. 
\item 
the ALICE collaboration obtained the multiplicities $dn_{\bar{\Lambda}}/dy|_{y=0} = 0.047 \pm 0.0053$ \cite{Aamodt:2011zza} and $dn_{\bar{p}}^0/dy|_{y=0} = 0.079\pm 0.0063$ at $\sqrt{s}=900\gev$~\cite{Aamodt:2011zj}. However, the hyperon multiplicity was corrected for feed-down from $\bar{\Xi}$ decay, while we are interested in all $\bar{\Lambda}$. In order to undo this unwanted feed-down correction we have to multiply $dn_{\bar{\Lambda}}/dy|_{y=0}$ by a factor of 1.12~\cite{Aamodt:2011zza}. We thus find $\bar{\Lambda}/\bar{p} =0.67\pm 0.09$.
\item
the CMS collaboration provides $dn_{\bar{\Lambda}+\Lambda}/dy|_{y=0} = 0.108\pm 0.012,\; 0.189 \pm 0.022$ \cite{Khachatryan:2011tm} and $dn_{\bar{p}}^0/dy|_{y=0} = 0.104\pm 0.004,\; 0.180\pm 0.0067$ at $\sqrt{s}=900\gev,\; 7000\gev$~\cite{Chatrchyan:2012qb}. At these ultra-high energies the production of $\bar{\Lambda}$ and $\Lambda$ is symmetric such that $dn_{\bar{\Lambda}}/dy|_{y=0} = 0.5\:dn_{\bar{\Lambda}+\Lambda}/dy|_{y=0}$~\cite{Aamodt:2011zza}. Again we have to apply the correction factor 1.12 to undo the feed-down from $\bar{\Xi}$ decay. We arrive at $\bar{\Lambda}/\bar{p} =0.58\pm 0.07,\; 0.59\pm 0.07$ for the two energies.
\end{itemize}
The data are collected in figure~\ref{fig:hyperon}. Despite significant scatter one can infer that the ratio remains nearly constant at $\bar{\Lambda}/\bar{p}\sim 0.3$ for $\sqrt{s} \lesssim 50\gev$ (this also corresponds to the value used in~\cite{Kappl:2014hha}). At $\sqrt{s}\gtrsim 50\gev$ $\bar{\Lambda}/\bar{p}$ increases notably before turning into a plateau at LHC energies. It is noteworthy that a similar increase followed by a plateau has been observed in the charged $K/\pi$ ratio (see e.g. figure 16 in~\cite{Adam:2015qaa}). 

One may speculate about the origin of the enhanced strangeness production. Within the Regge theory~\cite{Regge:1959mz}, soft hadronic processes at low $\sqrt{s}$ are dominated by pion exchange, while pomeron (glueball) exchange dominates at large $\sqrt{s}$. The increase of $\bar{\Lambda}/\bar{p}$ may then be related to the transition between the two regimes. We parameterize
\begin{equation}\label{eq:lambdapara}
\bar{\Lambda}/\bar{p}= c_1 +  \frac{c_2}{1+(c_3/s)^{c_4}}\,.  
\end{equation}
This form was chosen to match the observation of a constant ratio at low and at high energy. The parameters $c_3$ and $c_4$ determine at which energy and how rapidly $\bar{\Lambda}/\bar{p}$ increases. The best fit is obtained for $c_1=0.31$, $c_2=0.30$, $c_3=(146\gev)^2$, $c_4=0.9$. For determining the uncertainty, we randomly generated a large sample of tuples \{$c_1$, $c_2$, $c_3$, $c_4$\} according to their likelihood which we defined through a $\Delta\chi^2$ metric.\footnote{This was done by generating $\chi^2$ random variates. Each random variate is interpreted as a value of $\Delta\chi^2$ which defines a hyper-surface in the parameter space spanned by the $c_i$. One combination \{$c_1$, $c_2$, $c_{3}$, $c_4$\} on the hyper-surface is randomly picked and this procedure is repeated for every random variate.} In order to avoid unphysical energy-dependence we, furthermore, restrict the parameter range to $c_4=0.5-1.5$.\footnote{Due to the sparseness of data at intermediate energies the parameter $c_4$ is not well constrained. In particular very large values of $c_4$ for which $\bar{\Lambda}/\bar{p}$ approaches a step function in energy cannot be excluded on statistical grounds. We reject such unphysical behavior by limiting $c_4$ to the range indicated above.} At each energy, the half-width of the band shown in figure~\ref{fig:hyperon} corresponds to the standard deviation of the predicted $\bar{\Lambda}/\bar{p}$ ratio within the sample of parameter tuples. The fraction of hyperon-induced antiprotons is then determined by~\eqref{eq:convert}.

\section{Antiproton Multiplicity}\label{sec:multiplicity}

Measured antiproton multiplicities provide an important test for the parameterizations of  antiproton production employed in cosmic ray studies. The standard reference for antiproton multiplicities is still the work of Antinucci et al.~\cite{Antinucci:1972ib} which covers experimental data of the 1960s and early 1970s. As the energy range is limited to $\sqrt{s}\leq  53\gev$ it appears desirable to extend~\cite{Antinucci:1972ib} to higher energies relevant for present cosmic ray experiments. For this purpose we make use of collider data collected at RHIC and LHC.

Several complications occur: first, high energy data only exist on $dn_{\bar{p}}/dy|_{y\simeq 0}$, the multiplicity at central rapidity. A reliable extrapolation into the remaining phase space is required. Second, the values of $dn_{\bar{p}}/dy|_{y\simeq 0}$ published by different experiments do not necessarily refer to the same quantity. In some cases a feed-down correction has been applied, in others not. In addition, some experiments apply cuts to their scattering events which affect the resulting multiplicity.

Inelastic hadronic scatterings are divided into non-diffractive, single-diffractive and double-diffractive processes. Collisions are called diffractive if no internal quantum numbers are (and only little energy is) exchanged between the colliding particles. One distinguishes between single- and double-diffraction depending on whether one or both protons dissociate into multi-particle final states (in single-diffraction one proton remains intact). In both cases, there is no (or little) hadronic activity in the phase space region far from the the initial protons, i.e.\ there appears a rapidity gap. Non-diffractive processes do not fall into this category and show a spread of final state hadrons over the whole phase space. For the purpose of background reduction, some experiments require coincident particle detection in opposite pseudorapidity hemispheres. Such a selection rejects or at least reduces single-diffractive scatterings. The published antiproton multiplicity is then defined as the number of antiprotons per event passing this selection. As we are interested in the multiplicity per inelastic event, we have to apply a correction factor to the experimental result which accounts for the difference. Multiplicities are in general higher in non-diffractive events compared to diffractive events and, hence, this correction factor is always smaller than unity. We will now introduce the data sets included in our analysis. The first two were taken at RHIC, the others at LHC. Whenever a correction factor related to the event selection is applied we mention it explicitly.
\begin{itemize}
  \item the PHENIX collaboration determined the antiproton multiplicity at central rapidity with and without feed-down correction~\cite{Adare:2011vy}. Since a spurious effect occurs in the feed-down subtraction performed in~\cite{Adare:2011vy} -- in some momentum bins the antiproton production is reduced by a factor of three which appears suspiciously large -- we decided to use the data without feed-down correction to find\footnote{In~\cite{Adare:2011vy} this quantity is only given for the feed-down corrected case. We used the binned cross section data to arrive at the multiplicities for the uncorrected case.} $dn_{\bar{p}}/dy|_{y\simeq 0} = 0.036\pm 0.004,\: 0.062\pm 0.007$ at $\sqrt{s}=62\gev,\:200\gev$.     
  \item the STAR collaboration obtained $dn_{\bar{p}}/dy|_{y\simeq 0} = 0.113\pm 0.010$ at $\sqrt{s}=200\gev$~\cite{Abelev:2008ab}. The data were not feed-down corrected. Furthermore, they refer to a non-single-diffractive selection, i.e.\ a correction factor must be applied to normalize the multiplicity to inelastic events.
  \item the ALICE collaboration found $dn_{\bar{p}}^0/dy|_{y\simeq 0} = 0.079\pm 0.008$ at $\sqrt{s}=900\gev$~\cite{Aamodt:2011zj}. With the superscript $0$ we indicated that feed-down subtraction was performed.
    \item the CMS collaboration determined $dn_{\bar{p}}^0/dy|_{y\simeq 0} = 0.104\pm 0.004,\:0.133\pm 0.005$ at $\sqrt{s}=900\gev,\: 2760\gev$ (again including feed-down correction)~\cite{Chatrchyan:2012qb}. In this case, however, a correction factor must be applied as a so-called double-sided selection was employed. The latter is similar but not equal to a non-single-diffractive selection as it rejects nearly all single-diffractive events, but also reduces the efficiency for double-diffractive and (to lesser extent) non-diffractive events.
\end{itemize}
The calculation of the multiplicities $n_{\bar{p}}$ than proceeds as follows: at each considered energy we generate a large sample of inelastic proton proton scattering events with PYTHIA (version 8.2)~\cite{Sjostrand:2006za,Sjostrand:2014zea}. From the sample, we extract the total number of antiprotons $N_{\bar{p},\text{tot}}$ over the whole phase space and the number of antiprotons $N_{\bar{p},|y|<0.5}$ in the rapidity window $|y|<0.5$. The total multiplicity is then given as $n_{\bar{p}} = (N_{\bar{p},\text{tot}}/N_{\bar{p},|y|<0.5})\times dn_{\bar{p}}/dy|_{y\simeq 0}$. In the case of STAR and CMS we additionally apply the correction factor related to the event selection. The latter is determined from the event sample by employing the same cuts as have been used in the experimental analysis. Let us note that -- depending on whether a feed-down correction had been applied to the data -- we either arrive at $n_{\bar{p}}^0$ or $n_{\bar{p}}$. Both quantities can be converted into each other as described in section~\ref{sec:hyperon}.

Finally we need to determine the systematic error of our procedure. The output of PYTHIA is subject to uncertainties following from the incomplete understanding of hadronization and multiparton interactions. Several parameters which can be set by the user as well as the choice of parton distribution functions affect the multiplicity significantly. We have, therefore, determined $n_{\bar{p}}$ separately for the 32 implemented PYTHIA tunes which roughly capture the allowed range of parameter variations. For each tune we additionally varied the relative strength of non-, single- and double-diffraction within experimental uncertainties\footnote{We vary the relative contributions of single- and double-diffraction to the total inelastic cross section in the range $15-25\%$ and $5-15\%$ respectively. This corresponds to the range of uncertainty suggested by the data~\cite{Armitage:1981zp,Ansorge:1986xq,Bernard:1986yh,Abelev:2012sea}.}. The systematic error on on the multiplicity is then taken to be half the difference between the maximal and minimal $n_{\bar{p}}$ within the 32 tunes and within the considered uncertainty of diffraction. 

\begin{table}[htp]
\begin{center}
\begin{tabular}{|cccc|}
\hline 
\rowcolor{light-gray}&&&\\[-3mm]
\rowcolor{light-gray} $\sqrt{s}$~[GeV] & $n_{\bar{p}}^0$ & $n_{\bar{p}}$ & Experiment\\[1mm]
\hline
&&&\\[-3mm]
$62$ & $\,0.109 \pm 0.015\,$  & $\,0.141 \pm 0.019\,$ & PHENIX~\cite{Adare:2011vy} \\[1mm]
$200$ & $\,0.255 \pm 0.035\,$  & $\,0.358 \pm 0.047\,$ & PHENIX~\cite{Adare:2011vy} \\[1mm]
$200$ & $\,0.397 \pm 0.061\,$  & $\,0.557 \pm 0.084\,$ & STAR~\cite{Abelev:2008ab} \\[1mm]
$900$ & $\,0.606 \pm 0.067\,$  & $\,0.898 \pm 0.103\,$ & ALICE~\cite{Aamodt:2011zj} \\[1mm]
$900$ & $\,0.647 \pm 0.066\,$  & $\,0.958 \pm 0.102\,$ & CMS~\cite{Chatrchyan:2012qb} \\[1mm]
$2760$ & $\,0.956 \pm 0.102\,$  & $\,1.426 \pm 0.158\,$ & CMS~\cite{Chatrchyan:2012qb} \\ \hline
\end{tabular}
\end{center}
\caption{Antiproton multiplicity in proton proton scattering as derived in this study. We distinguish the multiplicity of promptly produced antiprotons ($n_{\bar{p}}^0$) and the multiplicity including antiprotons from hyperon decays ($n_{\bar{p}}$).}
\label{tab:multiplicity}
\end{table}

In table~\ref{tab:multiplicity} we present our results for the antiproton multiplicity one time excluding and one time including antiprotons from hyperon decay. 
We will now turn to the parameterization of antiproton production and test it against the found multiplicities in section~\ref{sec:violation}.

\section{Antiproton Production at low and intermediate energies}\label{sec:scaling}

It is convenient to express the invariant (prompt) antiproton production cross section as a function of the transverse momentum and the radial scaling variable. Written in this form, the invariant cross section has a radial scaling window at $10\gev \lesssim \sqrt{s} \lesssim 50 \gev$ in which it becomes independent of the center-of-mass energy~\cite{Feynman:1969ej,Taylor:1975tm,Tan:1982nc}:
\begin{equation}
f_{\bar{p}}^0(\sqrt{s},x_R,\pt) \longrightarrow f_{\bar{p}}^0(x_R,\pt)\,.
\end{equation}
In~\cite{Kappl:2014hha} precision data of the NA49 experiment~\cite{Anticic:2009wd} were used and it was found that -- to very good approximation -- the cross section can be parameterized as
\begin{equation}\label{eq:scalingcross}
f_{\bar{p}}^0 = R\:c_5 \,(1-x_R)^{c_6} \,\exp{ \left[-\frac{\mt}{c_7}\right] }\,,
\end{equation}
where the transverse antiproton mass $\mt=\sqrt{\pt^2 + m_p^2}$ was introduced. The best fit parameters were determined as $c_5 = 399\:\text{mb}\gev^{-2}$, $c_6=7.76$, $c_7=0.168\gev$. We take into account the (correlated) uncertainties on these parameters as well as a $6\%$ systematic uncertainty in the overall normalization~\cite{Anticic:2009wd}.

\begin{figure}[htp]
\begin{center}   
 \includegraphics[width=15cm]{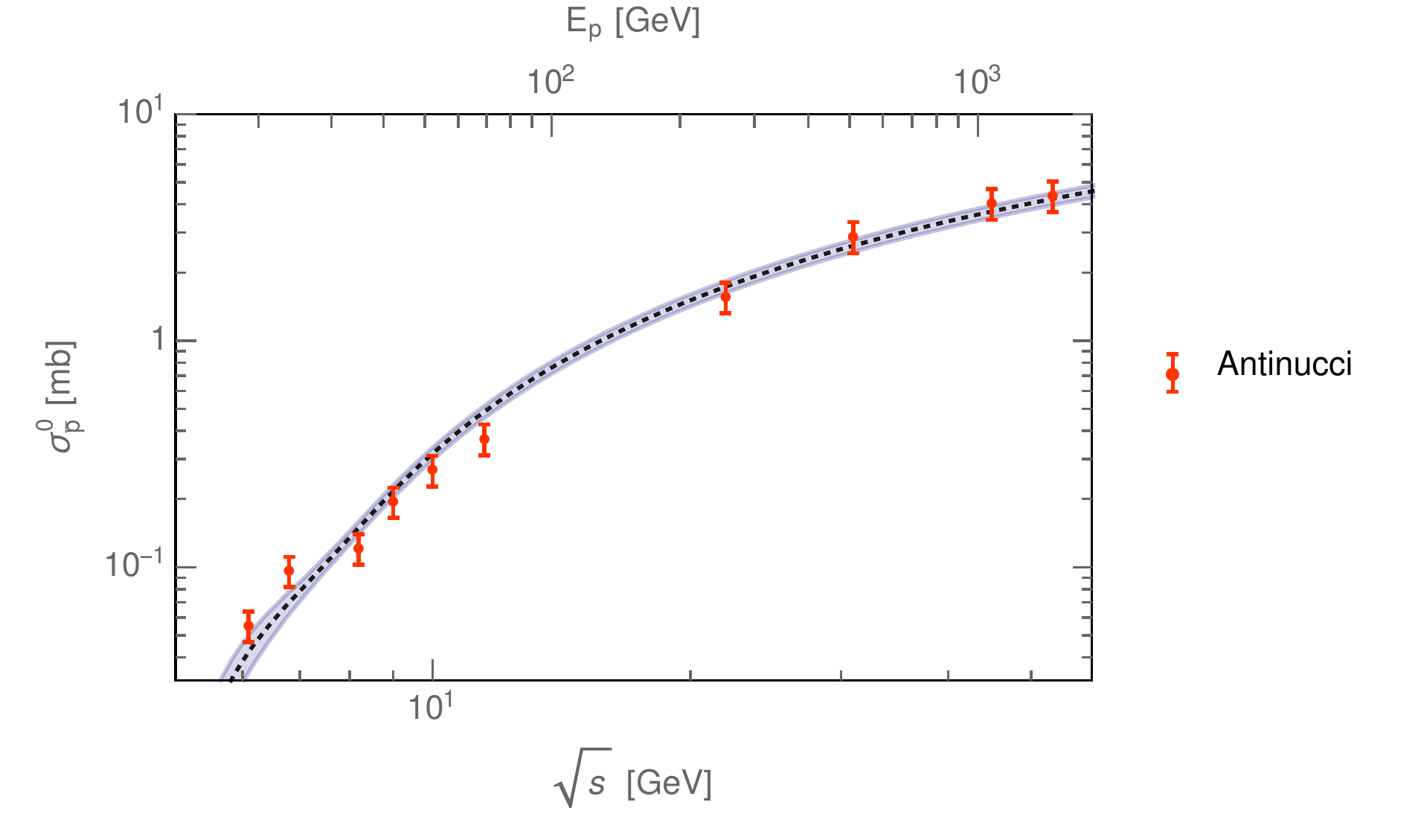}
\end{center}
\caption{Prompt antiproton production cross section (dashed line) and error band calculated from~\eqref{eq:scalingcross},~\eqref{eq:lowcorrection} compared with data of Antinucci et al.\ (error bars).}
\label{fig:scalingcrosssection}
\end{figure}

At energies below the scaling regime the correction factor $R$ needs to be included. The latter was determined in~\cite{Kappl:2014hha} by use of low energy accelerator data~\cite{Smith:1972ypc,Allaby:1970jt}. It was found that
\begin{align}\label{eq:lowcorrection}
 \!R= \begin{cases} \left[ 1 + c_9 \left( 10 - \tfrac{\sqrt{s}}{\text{GeV}}\right)^5\right] \exp\left[ c_{10} \left( 10 - \tfrac{\sqrt{s}}{\text{GeV}}\right) \left(x_R- x_{R,\text{min}}\right)^2\right]\,&\sqrt{s}\leq 10\gev\\ \;\;1  & \sqrt{s}> 10\gev\,,\end{cases}
\end{align}
where $x_{R,\text{min}}=m_p/E_{\text{max}}^*$. The parameters were determined as $c_9= (1\pm 0.4)\times 10^{-3}$ and $c_{10}= 0.7 \pm 0.04$.
In figure~\ref{fig:scalingcrosssection} we depict the integrated (prompt) antiproton production cross section calculated from~\eqref{eq:scalingcross},~\eqref{eq:lowcorrection}. For comparison we also show the cross sections determined by Antinucci et al.\ from experimental data at various energies~\cite{Antinucci:1972ib}.\footnote{Antinucci et al.\ provide the antiproton multiplicity $n_{\bar{p}}$. The latter was feed-down corrected (cf. section~\ref{sec:hyperon}) and multiplied by the inelastic cross section in order to arrive at the prompt antiproton production cross section $\sigma_{\bar{p}}^0$.}

The calculated cross section is in good agreement with the data of Antinucci et al.\ which confirms that radial scaling is valid up to energies $\sqrt{s}\sim 50\gev$.

\section{Scaling violation in the high energy regime}\label{sec:violation}

In this section we quantify the violation of radial scaling which occurs at high energies. When the scaling hypothesis was first formulated, Feynman assumed that $\sigma_{\text{in}}$ becomes constant at high energies~\cite{Feynman:1969ej}.

\begin{figure}[htp]
\begin{center}   
 \includegraphics[width=15cm]{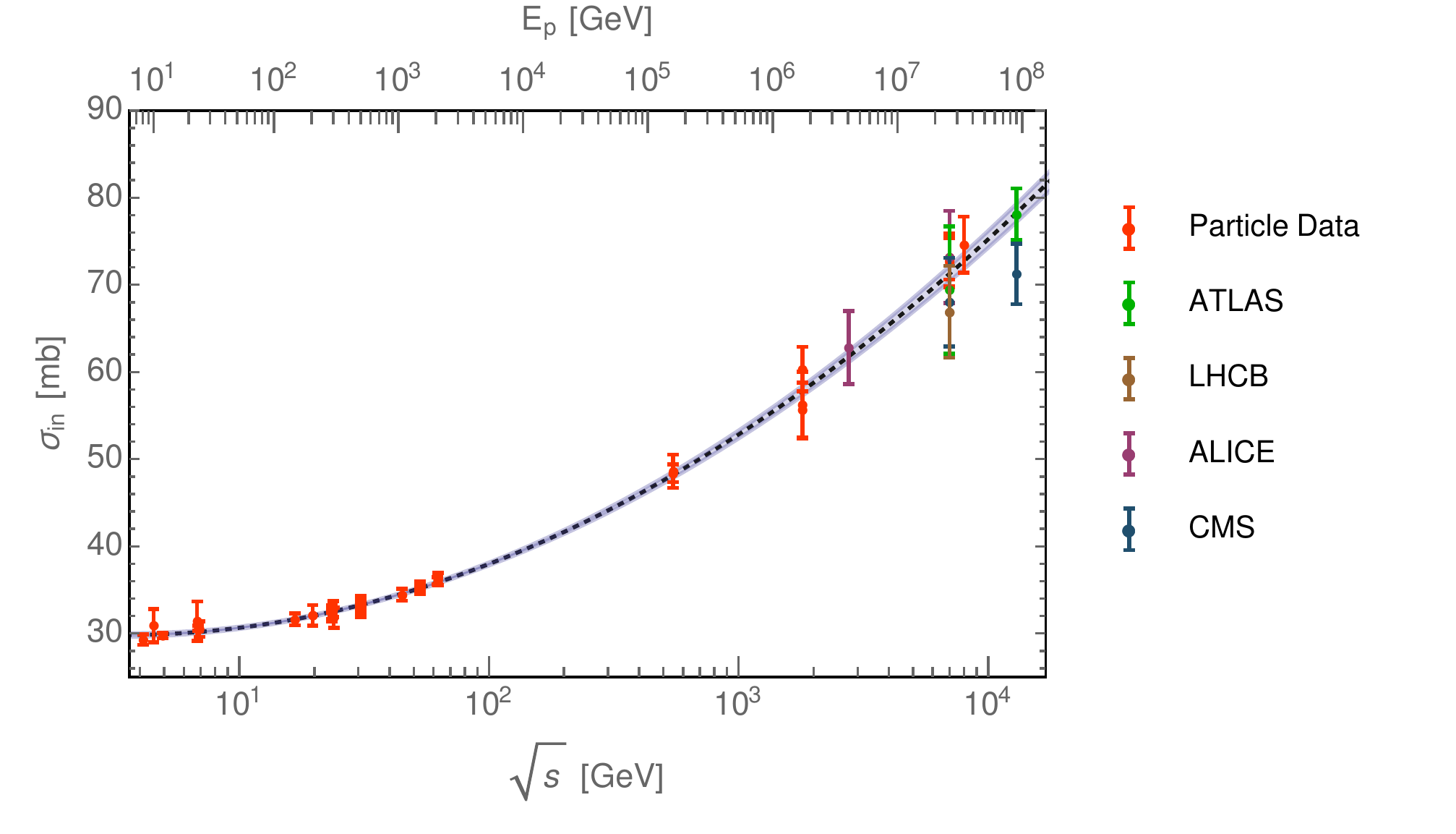}
\end{center}
\caption{Total inelastic cross section in proton proton scattering as measured by various experiments (error bars). The fit with the parameterization~\eqref{eq:inelasticpar} as well as the uncertainty band are also shown.}
\label{fig:inelastic}
\end{figure}

However, by now, a steady rise of $\sigma_{\text{in}}$ is established up to LHC energies. This can be seen in figure~\ref{fig:inelastic}, where we depict experimental data on the inelastic cross section in proton proton scattering collected by the particle data group~\cite{Olive:2016xmw} as well as new LHC measurements~\cite{Aad:2011eu,Zsigmond:2012vc,Abelev:2012sea,Aaij:2014vfa,Aaboud:2016mmw,VanHaevermaet:2016gnh}. We also include data from proton antiproton scattering~\cite{Olive:2016xmw} at $\sqrt{s}> 500 \gev$ as the particle or antiparticle nature of the initial state becomes irrelevant at such high energies. Also shown is our fit to the data using the parameterization suggested in~\cite{Block:2011vz}\footnote{Compared to~\cite{Block:2011vz} we neglect a term $\propto 1/\sqrt{s}$ as the coefficient in front of it is virtually vanishing at the best fit point.}
\begin{equation}\label{eq:inelasticpar}
 \sigma_{\text{in}} = c_{11} + c_{12}\, \log{\sqrt{s}} + c_{13}\, \log^2{\sqrt{s}}\,,
\end{equation}
where $\sqrt{s}$ enters in units of GeV. The best fit parameters are $c_{11}=30.9\mb$, $c_{12}=-1.74\mb$, $c_{13}=0.71\mb$ and the uncertainty band was determined by a $\Delta\chi^2$ test (as described in section~\ref{sec:hyperon}). The rise of the inelastic cross section suggests a modification of the radial scaling hypothesis (see~\cite{Kachelriess:2015wpa}). Instead of taking the invariant antiproton production cross section to be independent of $\sqrt{s}$, one might expect that it grows proportional to the inelastic cross section $f_{\bar{p}}^0 \propto \sigma_{\text{in}}$. 

\begin{figure}[t]
\begin{center}   
 \includegraphics[width=15cm]{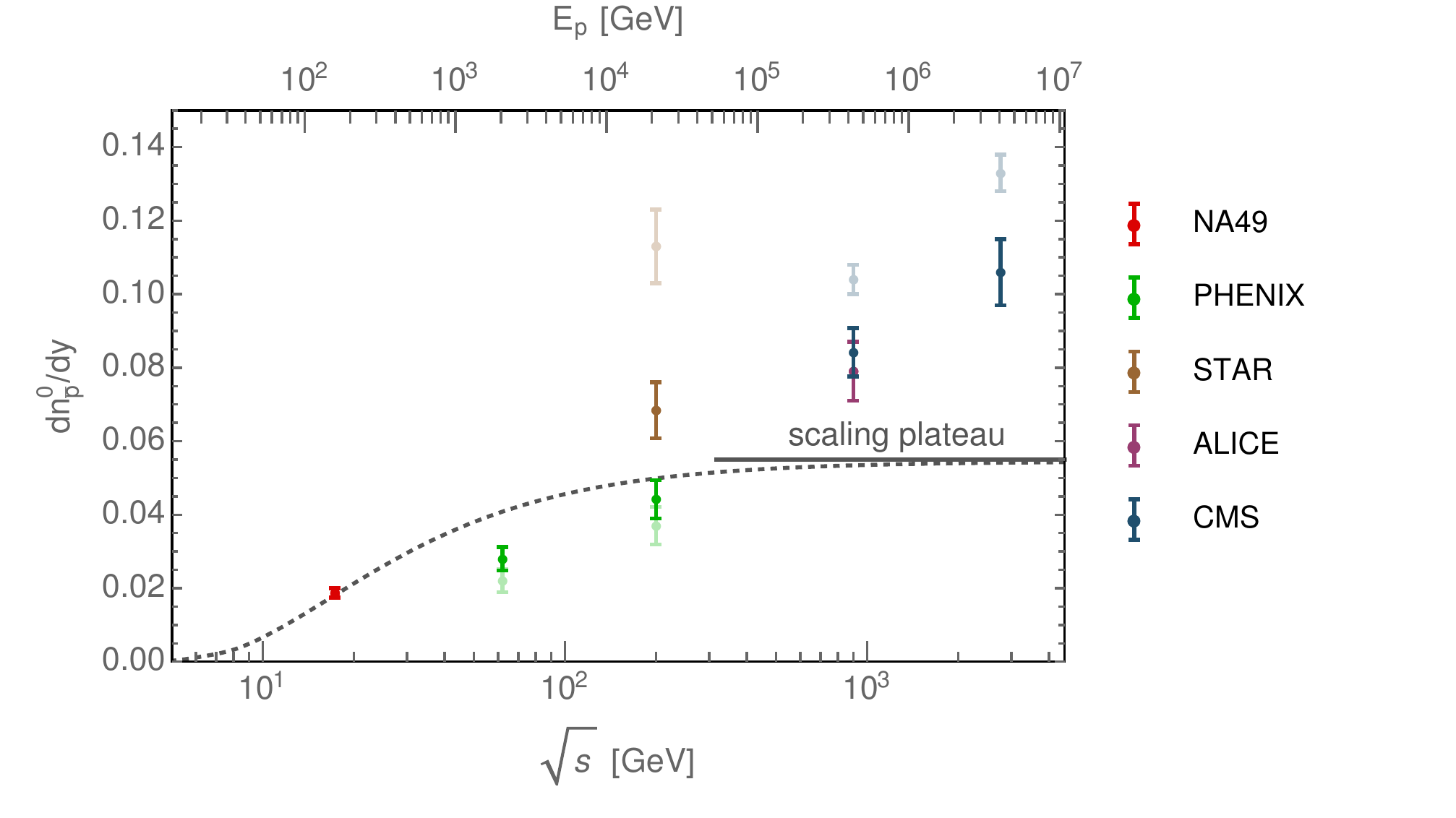}
\end{center}
\caption{Antiproton multiplicity at central rapidity as measured by various experiments. The light colored error bars refer to the published data, the full colored error bars are obtained after correcting for cuts and feed-down (see section~\ref{sec:multiplicity}). The dashed line turning into a plateau at high energies shows the theoretical prediction if radial scaling was preserved at the level of the multiplicity.}
\label{fig:centralrapidity}
\end{figure}

If the inelastic cross section was the only source of scaling violation, scaling would still be preserved at the level of the multiplicity, i.e.\ $E\, d^3 n_{\bar{p}}^0 /d^3 p$ would be independent of $\sqrt{s}$. Then, it would directly follow that the multiplicity at central rapidity ($y=0$) approaches a constant at high energies
\begin{equation}
 \left.\frac{dn_{\bar{p}}^0}{dy}\right|_{y=0} \;\,\xrightarrow{\sqrt{s}\:\rightarrow\:\infty}\;\; \text{constant}\,.
\end{equation}
However, as can be seen in figure~\ref{fig:centralrapidity}, experimental data from RHIC and LHC (see section~\ref{sec:multiplicity}) strongly suggest that such a plateau does not exist. Rather, $dn_{\bar{p}}^0/dy|_{y=0}$ keeps growing up to to the highest available energies. Hence, radial scaling is also broken for the multiplicity.

\begin{figure}[t]
\begin{center}   
 \includegraphics[width=15cm]{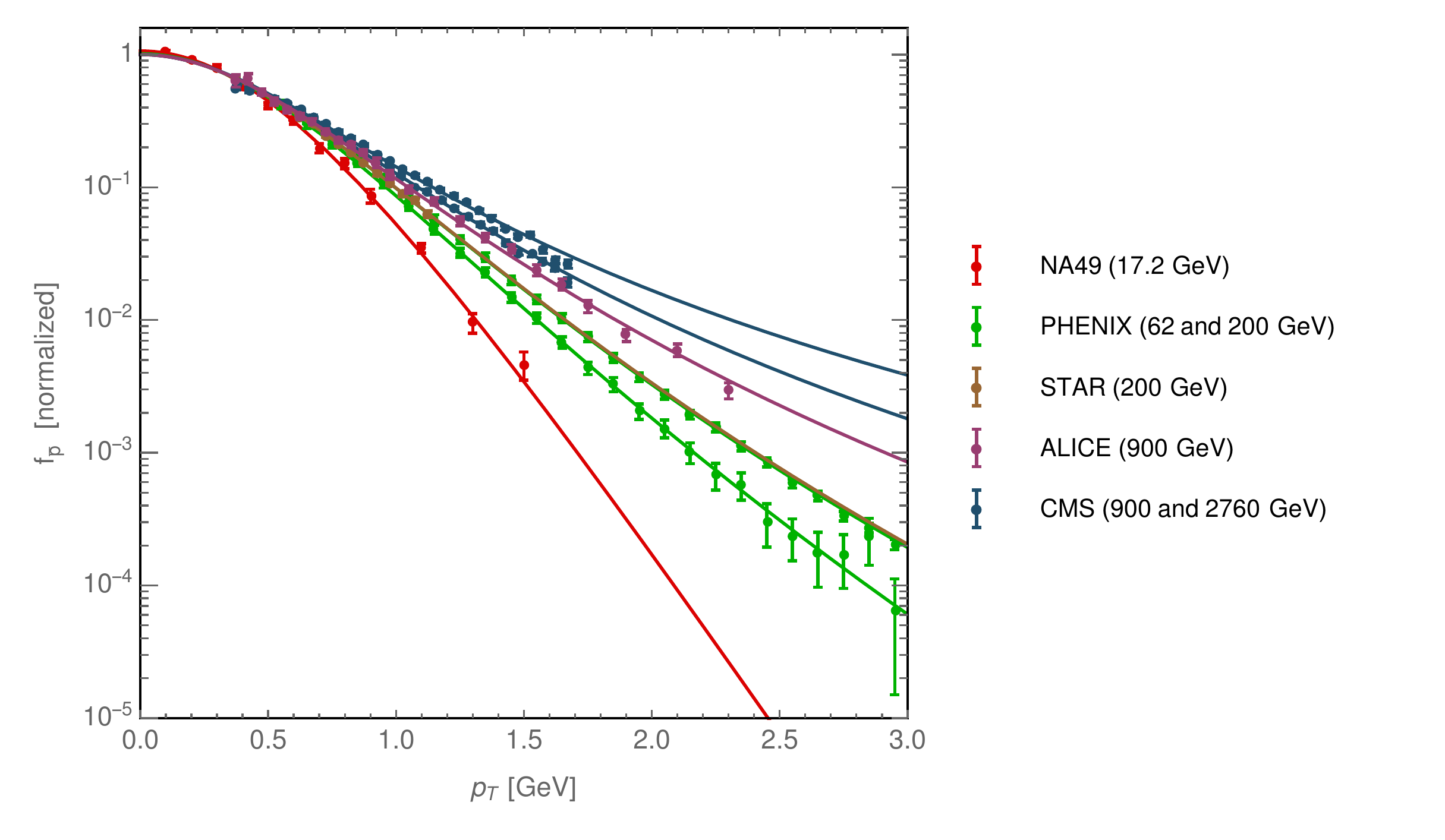}
\end{center}
\caption{Transverse momentum distribution of the invariant antiproton production cross section. Experimental data were taken at the center-of-mass energies indicated in the figure legend. For each data set, the normalization was chosen such that $f_{\bar{p}}=1$ at $\pt=0$. Colored lines indicate the fit using a Tsallis distribution (see text).}
\label{fig:pt}
\end{figure}

The additional source of scaling violation can be identified within the antiproton transverse momentum distributions. In figure~\ref{fig:pt} we depict $f_{\bar{p}}$ at central rapidity as a function of the transverse momentum. The data sets of NA49~\cite{Anticic:2009wd}, PHENIX~\cite{Adare:2011vy}, STAR~\cite{Abelev:2008ab}, ALICE~\cite{Aamodt:2011zj} and CMS~\cite{Chatrchyan:2012qb} cover center-of-mass energies from $\sqrt{s}=17.2\gev$ to $2760\gev$. In order to focus on the spectral distribution, all data sets were normalized such that $f_{\bar{p}}=1$ at $\pt=0$.

For NA49, the transverse mass distribution falls exponentially in $\mt$~\cite{Kappl:2014hha}. However, the other data sets indicate deviations from the exponential form which grow with collision energy. Indeed, at high transverse momentum, the spectra can be shown to follow a power law in $\mt$. The transition from the exponential to the power law moves to smaller $\pt$ with increasing collision energy. The same behavior of transverse momentum spectra has also been observed for pions, kaons, protons and other hadrons (see e.g.~\cite{Marques:2015mwa}). The spectra can be successfully fit with a Tsallis distribution~\cite{Tsallis:1987eu} which may suggest a thermodynamical interpretation of high energy hadron collisions~\cite{Azmi:2014dwa}. 

In order to account for the observed transverse momentum behavior, we modify the invariant antiproton production cross section in the following way
\begin{equation}\label{eq:tsallis}
 f_{\bar{p}}^0 = R \:\sigma_{\text{in}} \, c_5 \,(1-x_R)^{c_6} \,\big[1 + X (\mt - m_p)\big]^{-\frac{1}{X c_7}}\,.
\end{equation}
Compared to~\eqref{eq:scalingcross} this includes the dependence on $\sigma_\text{in}$ suggested by the rise of the inelastic cross section. Furthermore, the function $X$ was introduced. In the limit $X\rightarrow 0$ the exponential dependence on $\mt$ of~\eqref{eq:scalingcross} is recovered. However, we now take $X$ to be energy-dependent in order to model the increase of cross section at large $\pt$. Both, $\sigma_\text{in}$ and $X$, constitute sources of radial scaling violation. The parameters $c_5$, $c_6$ and $c_7$ and the corresponding uncertainties are again obtained by a fit to the NA49 data ($c_5=0.047\gev^{-2}$, $c_6=7.76$, $c_7=0.168\gev$ at the best fit point). In order to determine $X$ as a function of the collision energy, we fit the distribution~\eqref{eq:tsallis} with a free normalization to the experimental data of NA49, PHENIX, STAR, ALICE and CMS (see figure~\ref{fig:pt}). We find that $X$ can be parametrized as
\begin{equation}\label{eq:X}
 X = c_8\, \log^2\left[\frac{\sqrt{s}}{\sqrt{s}_{\text{th}}}\right]\,,
\end{equation}
where $c_8=(0.038\pm 0.00057)\gev^{-1}$ and $\sqrt{s}_{\text{th}}=4\,m_p$ denotes the threshold center-of-mass energy for antiproton production.

\begin{figure}[t]
\begin{center}   
 \includegraphics[width=15cm]{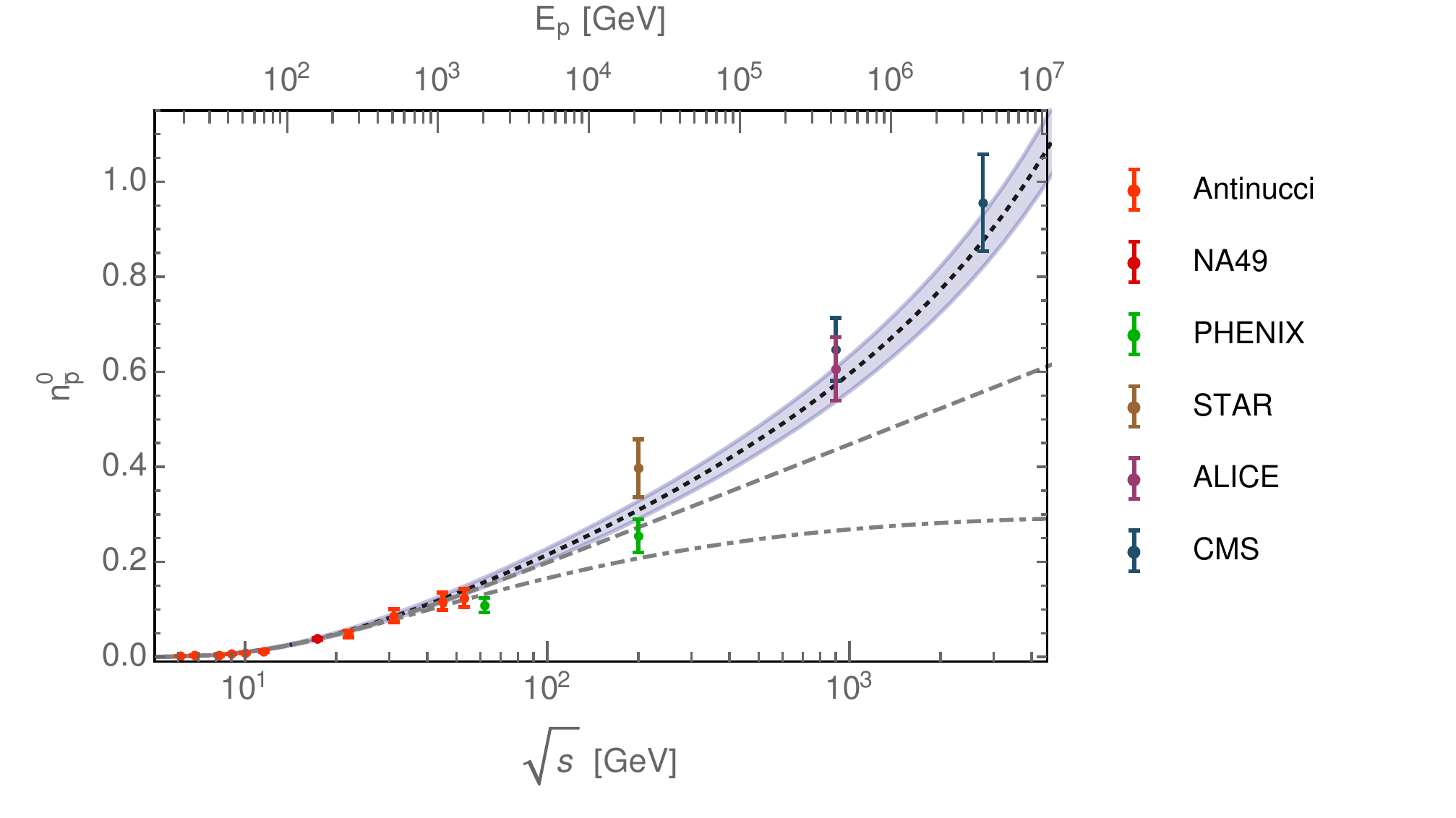}
\end{center}
\caption{Prompt antiproton multiplicity and uncertainty band predicted by the new cross section parameterization in this work (shaded band). Experimental data from Antinucci et al.~\cite{Antinucci:1972ib} and from table~\ref{tab:multiplicity} are shown for comparison. The dash-dotted gray line indicates the multiplicity obtained in the absence of high energy scaling violation. The dashed gray line includes scaling violation in the inelastic cross section, but not in the transverse momentum distribution.}
\label{fig:multiplic}
\end{figure}

In figure~\ref{fig:multiplic} we compare the prompt antiproton multiplicity as obtained before and after including the effects of scaling violation. The scaling-preserving parameterization~\eqref{eq:scalingcross} is in good agreement with the data up to energies $\sqrt{s} \sim 50\gev$. However, at higher energies, scaling violation becomes significant and~\eqref{eq:scalingcross} loses its validity. The new parameterization~\eqref{eq:tsallis}, on the other hand, reliably predicts the multiplicity over the full available energy range. This suggests that~\eqref{eq:scalingcross} now indeed contains all relevant sources of scaling violation which are relevant up to TeV energies.

\section{Comparison with Previous Work}\label{sec:previouswork}

We will now compare our determination of the antiproton production cross section with the existing literature. We consider proton proton scattering in the laboratory frame, where an incoming proton with kinetic energy $T'$ hits a proton at rest. We define
\begin{equation}
 \tilde{q} = \int dT' \;(T')^{-3} \;\frac{d\sigma_{\bar{p}}}{dT} \,,
\end{equation}
where $d\sigma_{\bar{p}}/dT$ denotes the differential antiproton production cross section and $T$ the kinetic energy of the produced antiproton. As the 
\begin{figure}[htp]
\begin{center} 
\includegraphics[width=12.5cm]{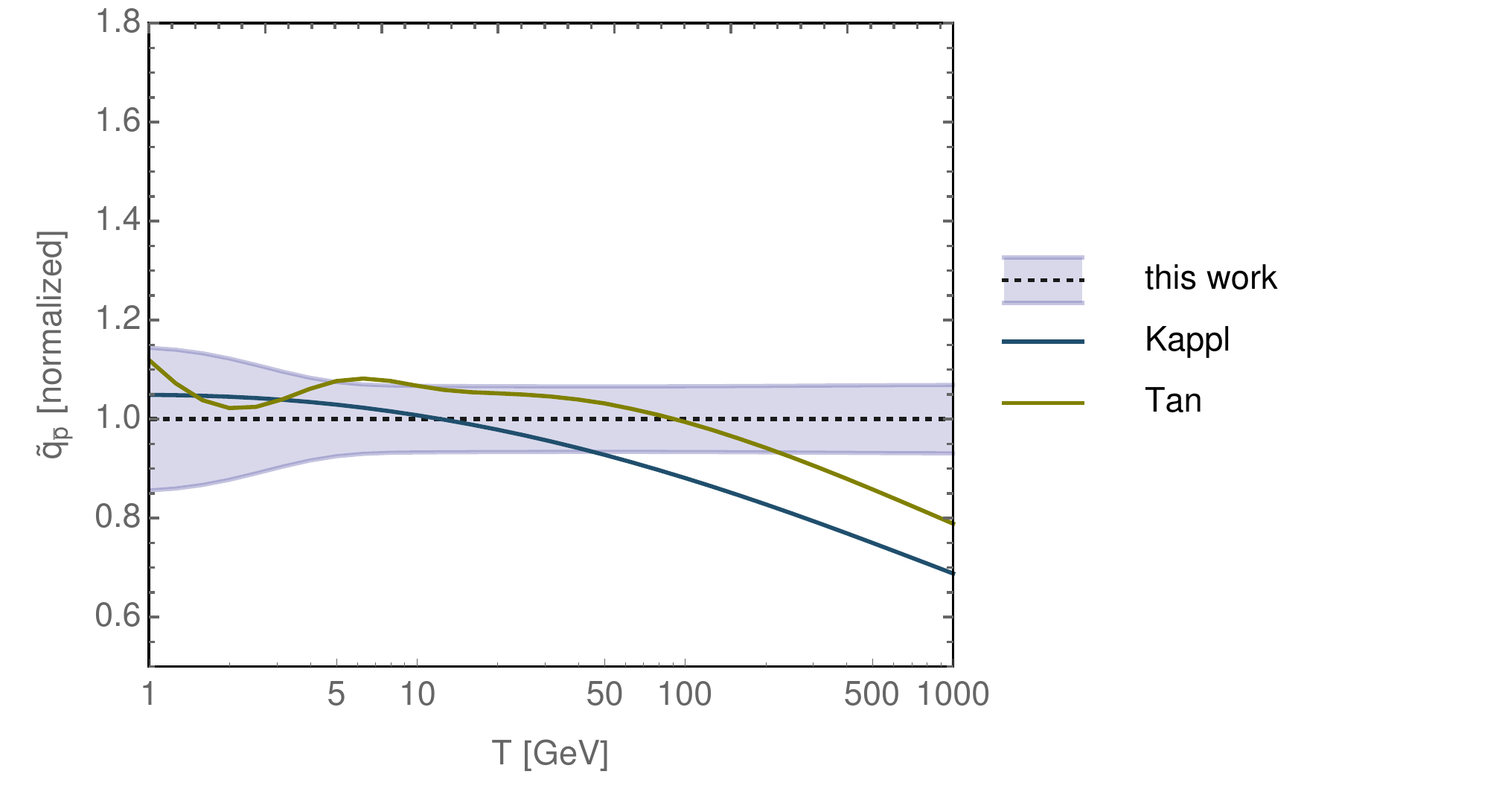}\\
\includegraphics[width=12.5cm]{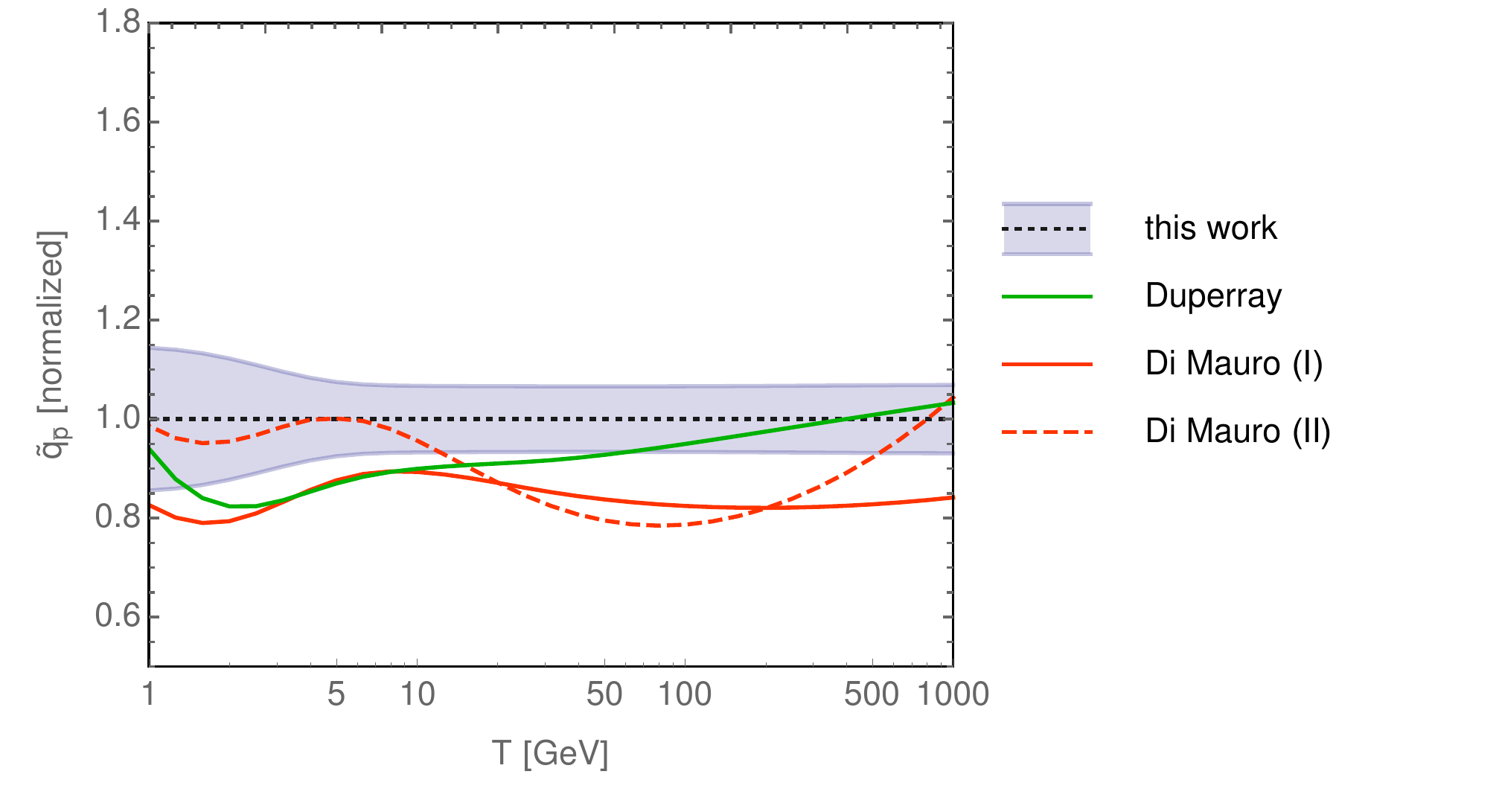}\\
\includegraphics[width=12.5cm]{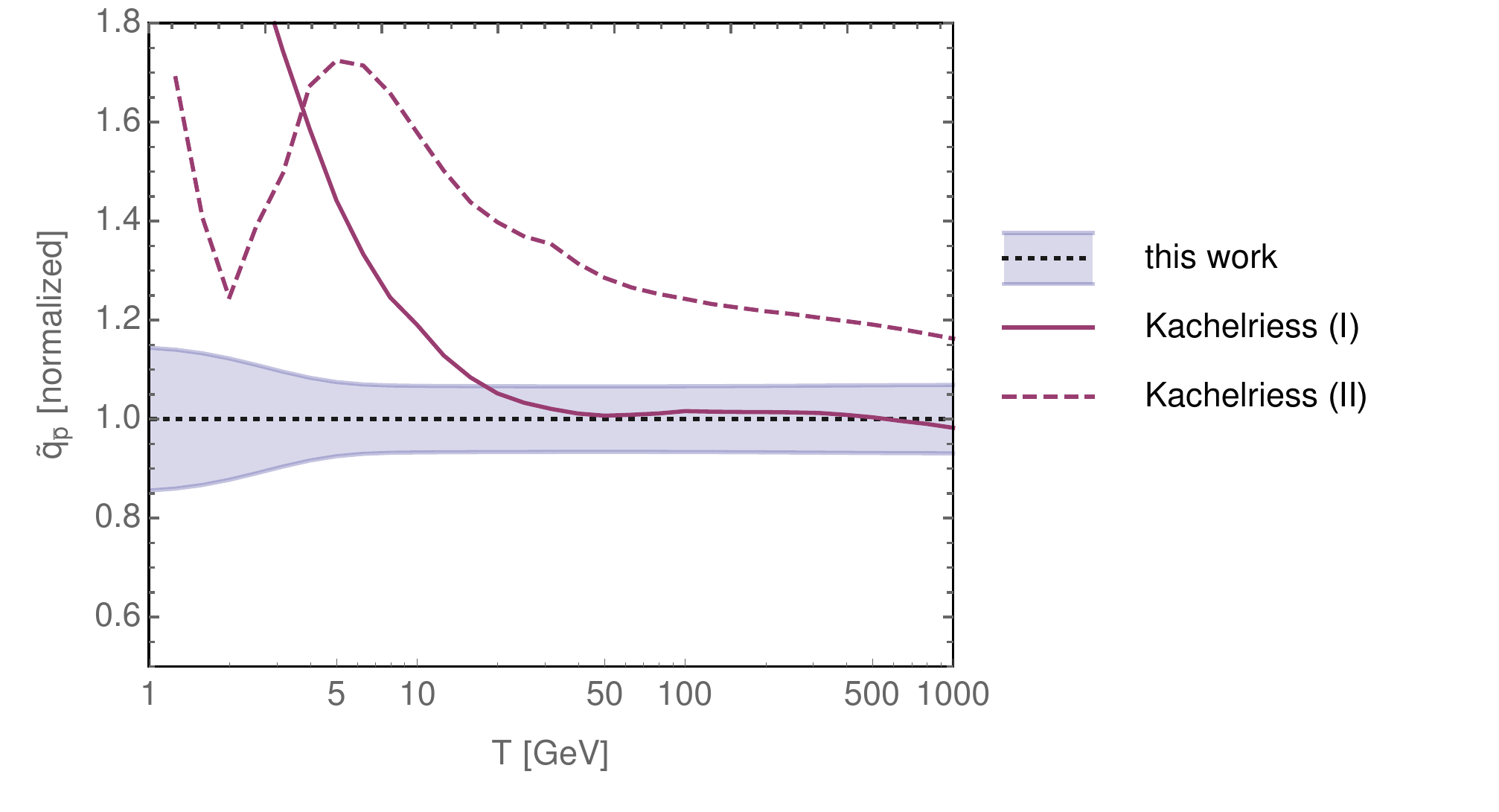}
\end{center}
\caption{Previous evaluations of the antiproton production cross section compared to this work (see text).}
\label{fig:comparison}
\end{figure}
cosmic ray proton flux falls of with a power law similar to $-3$, $\tilde{q}$ is closely related to the antiproton production rate in the galactic disc and, therefore, a useful quantity for our comparison.
We calculated $\tilde{q}$ for our previous parameterization from 2014~\cite{Kappl:2014hha}, for the parameterizations of Tan and Ng (1982)~\cite{Tan:1982nc}, Duperray et al.\ (2003)~\cite{Duperray:2003bd}, di Mauro et al.\ (2014)~\cite{diMauro:2014zea} as well as for the Monte Carlo based evaluation of Kachelriess et al.\ (2015)~\cite{Kachelriess:2015wpa}. In figure~\ref{fig:comparison} we depict the different $\tilde{q}$ normalized to the one of our present study.

Despite the fact that Tan and Ng's parameterization was derived from old experimental data, it is still in reasonable agreement with the present work up to antiproton energies $T\sim 50\gev$. As in~\cite{Kappl:2014hha}, radial scaling was assumed leading to an underestimation of antiproton production at high energies. This is clearly visible in figure~\ref{fig:comparison}.

Di Mauro et al.\ give two possible parameterizations of the invariant antiproton cross section in equations (12) and (13) of~\cite{diMauro:2014zea} which will be dubbed (I) and (II). The parameterizations derive from that of Duperray et al., but use updated experimental data and contain modifications of the antiproton transverse momentum distribution. At antiproton energies $T < 100\,\gev$ (I) and (II) both predict cross sections which are systematically a bit lower than those derived in the present work. It is likely that this difference is caused by hyperon-induced antiprotons which have not been added to the NA49 data in~\cite{diMauro:2014zea}. Both parameterizations (I) and (II) of di Mauro et al.\ contain several explicit sources of scaling violation. The corresponding parameters were fit using experimental data covering the energy range $\sqrt{s} = 6 -200\gev$. At higher energy the extrapolation introduces uncertainties. This becomes manifest if one uses the parameterizations of di Mauro et al.\ to predict the antiproton multiplicity at $\sqrt{s} =900\gev$. For (I) a value of $n_{\bar{p}}=0.90$ consistent with the ALICE and CMS data (cf. table~\ref{tab:multiplicity}) is obtained. Parameterization (II), however, predicts $n_{\bar{p}}=16.9$ which is too high by a factor of $\sim 20$. Correspondingly, a reasonable high energy behavior of $\tilde{q}$ is found for parameterization (I), while the fast rise of $\tilde{q}$ at $T>100\gev$ predicted by (II) is inconsistent with experimental data.

The predictions obtained by Kachelriess et al.\ with the Monte Carlo generators QGSJET-IIm (I) and EPOS-LHC (II) provide a very useful cross-check in the high energy regime. As can be seen in figure~\ref{fig:comparison}, in particular the $\tilde{q}$ obtained with QGSJET-IIm agrees very well with our $\tilde{q}$ for $T\gtrsim 50\gev$. Significant discrepancies occur at low energies. These can be traced back to the breakdown of hadronization models implemented in Monte Carlo generators. Very similar observations had already been made for the Monte Carlo generators PYTHIA, DPMJET and GEANT~\cite{Kappl:2014hha}.

\section{Isospin Effects}\label{sec:isospin}

About half of the antiprotons in cosmic rays stem from the decay of long-lived antineutrons. Due to the lack of experimental data on antineutron production this contribution can only be estimated from symmetry arguments. While in older cosmic ray studies equal production of antiprotons and antineutrons was assumed~\cite{Tan:1982nc,Duperray:2003bd}, a possible asymmetry was considered in~\cite{Kappl:2014hha,diMauro:2014zea}. 
In~\cite{Fischer:2003xh,Chvala:2003dn} it was argued that such an asymmetry follows from underlying isospin effects. Proton and neutron can be identified as doublet under an SU(2) isospin group. Baryon number conservation dictates pair production of baryons, either as $\bar{p}n$, $\bar{p}p$, $\bar{n}n$ or $\bar{n}p$.\footnote{Production of pairs involving heavier baryons is of course also viable.} The asymmetric pairs $\bar{p}n$ and $\bar{n}p$ carry opposite isospin. Quantum number conservation in the microscopic processes may then lead to a preferred production of either $\bar{p}n$ or $\bar{n}p$ depending on the isospin of the colliding particles. In~\cite{Fischer:2003xh} proton proton and neutron proton scattering data at a collision energy $\sqrt{s}=17.2\gev$ were compared. It was found a higher antiproton multiplicity with neutron compared to proton projectiles. This suggests that -- at this energy -- $\bar{n}p$ final states are preferred over $\bar{p}n$ final states in the case of proton projectiles and vice versa for neutron projectiles. On the other hand it was argued in~\cite{Videbaek:1995mf} that isospin effects disappear at higher energies due to very efficient charge exchange reactions which interconvert protons and neutrons. This argument is supported by measurements of the antiproton-to-proton ratio at mid-rapidity. The latter is observed to approach unity in proton proton collisions at LHC energies implying equal production of $\bar{p}n$ and $\bar{n}p$ pairs.

In the following we will constrain possible isospin effects in proton proton scattering by the combination of experimental data and symmetry arguments. The isospin factor $\iso = f_{\bar{n}}^0 / f_{\bar{p}}^0 -1$ (cf.~\eqref{eq:contributions}) measures the enhancement of antineutron over antiproton production. In~\cite{Kappl:2014hha}, we conservatively estimated this factor to stay in the range $\iso= 0-0.43$.\footnote{Another estimate of the isospin factor $\iso=0.3\pm 0.2$ can be found in~\cite{diMauro:2014zea}.} In this work we go one step further and estimate the energy-dependence of the isospin factor. We take $\iso$ to be a function of $\sqrt{s}$, but otherwise phase-space independent.

First we employ neutron proton as well as proton nucleus scattering data. Following~\cite{Baatar:2012fua,Kappl:2014hha}, we express the differential antiproton multiplicity for scattering of the projectile $i$ on the target $j$ by the differential multiplicity in proton proton scattering\footnote{This equation is only applicable to scattering of nucleons or very light nuclei for which nuclear medium effects like the Cronin effect~\cite{Cronin:1973fd} can be neglected.}
\begin{equation}\label{eq:decomposition}
\left(\frac{dn_{\bar{p}}^0}{dx_f}\right)_{\!ij}   = \Big(\langle \nu_i \rangle \left(1+ \tfrac{N_i}{A_i}\,\iso\right) F_\text{pro}(x_f)
+\langle \nu_j \rangle \left(1+ \tfrac{N_j}{A_j}\,\iso\right) F_\text{tar}(x_f)\Big)
\left(\frac{dn_{\bar{p}}^0}{dx_f}\right)_{\!pp}\,.
\end{equation}
Here, the multiplicity was decomposed into a projectile and a target component. The projectile and target overlap functions $F_\text{pro}$ and $F_\text{tar}= 1 - F_\text{pro}$ were defined in~\cite{Baatar:2012fua} as functions of the Feynman scaling variable. In the far-forward hemisphere ($x_f\gg0$) only projectile factorization contributes to the multiplicity such that $F_\text{pro}=1$, while $F_\text{tar}=1$ in the far-backward hemisphere ($x_f\ll0$). However, projectile and target fragmentation slightly leak into the ``wrong'' hemisphere which implies that both $F_\text{pro}$ and $F_\text{tar}$ are non-zero around $x_F=0$. We, furthermore, defined $\langle \nu_{i,j} \rangle$ as the average number of interacting nucleons in the projectile and target respectively. For nucleons, deuterons, helium and carbon nuclei $\langle\nu_{p,n}\rangle=1$, $\langle\nu_{\text{D}}\rangle=1.05$, $\langle\nu_{\text{He}}\rangle=1.25$ and $\langle\nu_{\text{C}}\rangle=1.6$ respectively~\cite{Baatar:2012fua,Kappl:2014hha}. The neutron to nucleon numbers of the colliding particles $N_{i}/A_{i}$ and $N_{j}/A_{j}$ determine the strength of the isospin enhancement. We assumed that the antiproton multiplicity receives the same isospin enhancement in neutron scattering as the antineutron multiplicity in proton scattering. This assumption is justified by the the isospin symmetry.

The isospin enhancement can now be determined by fitting~\eqref{eq:decomposition} with free $\iso$ to a given set of scattering data. We include the following measurements:
\begin{itemize}
 \item NA49 measured the differential antiproton multiplicity in neutron proton scattering at $\sqrt{s}=17.2\gev$ in the window $x_f = -0.05 \dots 0.25$~\cite{Fischer:2003xh}. This data set was obtained from deuteron proton scattering by triggering on events with a spectator proton. The isospin factor follows from a comparison with the proton proton scattering data taken at the same energy~\cite{Anticic:2009wd}. Assuming a fully correlated systematic uncertainty\footnote{The correlated normalization uncertainty was estimated from the tabulated uncertainty in proton proton and proton carbon scattering~\cite{Baatar:2012fua}. If the normalization uncertainty was not taken into account $\iso=0.37\pm 0.06$ would be obtained~\cite{Kappl:2014hha}.} of $10\%$ in the relative normalization of both data sets, we find $\iso=0.13\pm 0.10$.
 \item In~\cite{Baatar:2012fua} the differential antiproton multiplicity in proton carbon scattering for $x_f=-0.2\dots 0.3$ and $\sqrt{s}=17.2$ was measured by NA49. By comparison with the proton proton data and taking into account the correlated normalization uncertainty of $10\%$, we obtain $\iso=0.17\substack{+0.08 \\ -0.22}$.
 \item Measurements of the invariant antiproton production cross section in proton proton and proton deuteron scattering were performed at Fermilab~\cite{Antreasyan:1978cw}. The data were collected at $\sqrt{s}=27.5\gev$ and are dominated by target factorization ($x_f \simeq -0.05$). Variations in the beam luminosity were estimated to induce a $10\%$ normalization uncertainty. Due to large statistical errors, the isospin factor is not well constrained. It is found that $\iso=0.458\substack{+0.88 \\ -0.50}$.
\end{itemize}
A second possibility to constrain isospin effects is to employ the $\bar{p}/p$ ratio in proton proton scattering. Isospin enhancement of the antineutron multiplicity is connected to a simultaneous excess of protons over antiprotons as it origins from the preferred production of $\bar{n}p$ pairs compared to $n\bar{p}$. In particular, it follows that $\bar{p}/p < (1 + \iso)^{-1}$. This constraint only becomes relevant at high energies where $\bar{p}/p$ approaches unity. From the measurements of $\bar{p}/p$ at mid-rapidity performed by ALICE~\cite{Aamodt:2010dx,Abbas:2013rua} we estimate that $\iso < 0.06$ at $\sqrt{s}=900\gev$, $\iso < 0.04$ at $\sqrt{s}=2760\gev$ and $\iso < 0.02$ at $\sqrt{s}=7000\gev$.

Finally, we compare the $\bar{p}/p$ ratio in proton proton and deuteron gold collisions observed by the STAR experiment at $\sqrt{s}=200\gev$~\cite{Abelev:2008ab}. It is found that $(\bar{p}/p)_{\text{D}\text{Au}}/(\bar{p}/p)_{pp}=1.0\pm0.1$, i.e.\ there is no indication of isospin effects (which would have enhanced $\bar{p}/p$ for nuclear compared to proton collisions). We conservatively estimate $\iso < 0.1$ at $\sqrt{s}=200\gev$.
\begin{figure}[htp]
\begin{center}   
 \includegraphics[width=15cm]{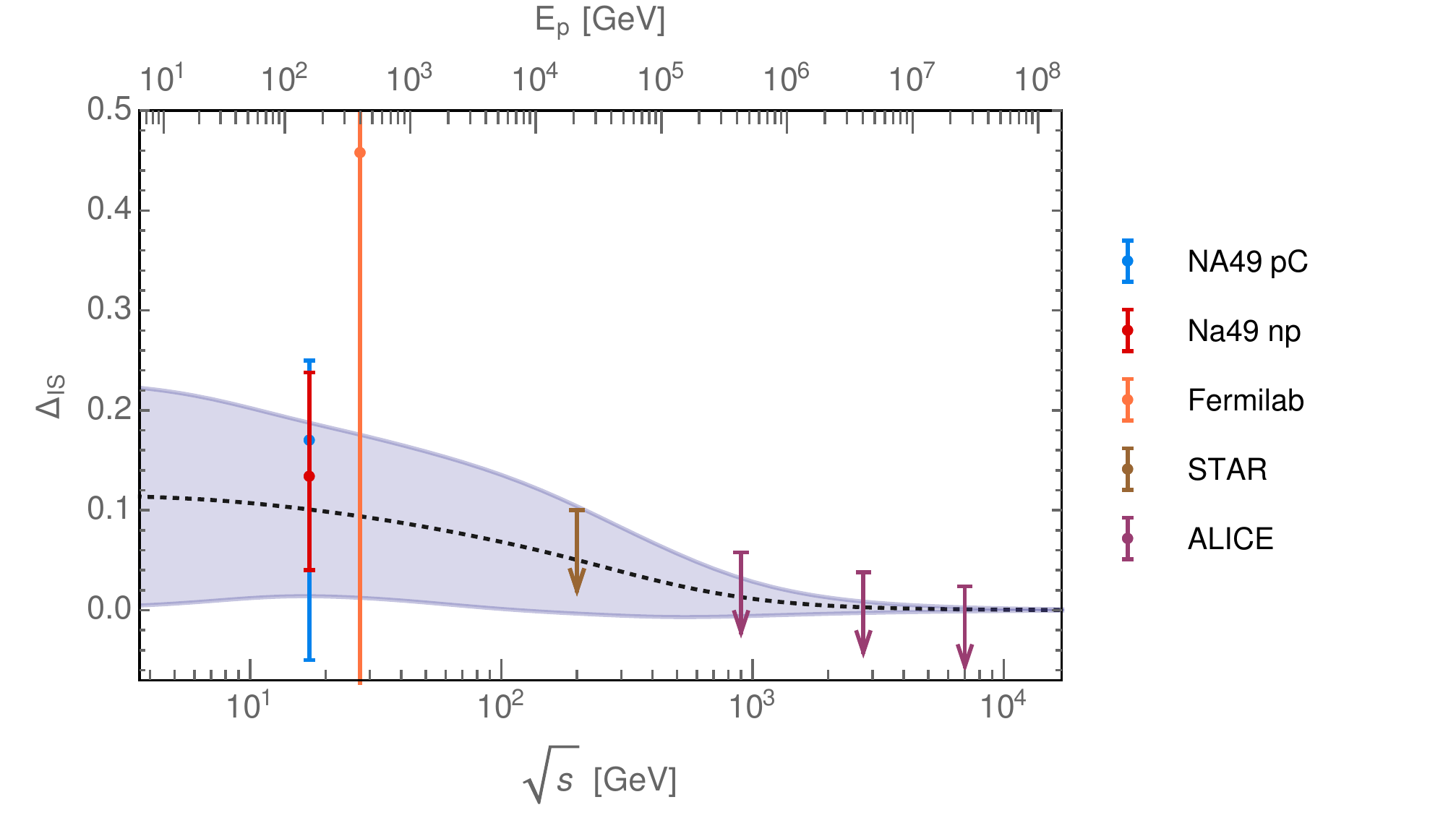}
\end{center}
\caption{Constraints on the isospin factor derived from experimental data (see text). Also shown is the fit using~\eqref{eq:isofit} and the uncertainty band.}
\label{fig:isospin}
\end{figure}

In figure~\ref{fig:isospin} we depict the constraints on $\iso$ derived above. While the low energy data favor a slight isospin enhancement of antineutron production, this enhancement disappears at high energy. As we mentioned previously, a similar effect appears for total inelastic cross sections which lose their sensitivity to the nature of the colliding hadrons at high energies. Both observations can be interpreted such that internal quantum number conservation affects the final state hadron production less with growing energy. In order to determine the energy-dependence of $\iso$ we perform a fit using the parameterization\footnote{As in~\eqref{eq:lambdapara} we restrict $n$ to the range $n=0.5-1.5$ in order to avoid unphysical energy-dependence.}
\begin{equation}\label{eq:isofit}
\iso =  \frac{c_{14}}{1+(s/c_{15})^{c_{16}}}\,,  
\end{equation}
which is chosen such that $\iso$ vanishes for $\sqrt{s} \rightarrow \infty$. The parameterization was fit to the experimental data, uncertainties were derived as in section~\ref{sec:hyperon}.

\section{Implications for the Antiproton Flux}\label{sec:flux}

Secondary antiprotons in the galaxy are produced by scattering of primary cosmic rays on the interstellar matter. They constitute the background in searches for annihilating dark matter or other exotic sources of antiprotons. The antiproton production cross sections enter the calculation of the secondary antiproton source term which is defined as the differential antiproton production rate per volume, time and energy
\begin{equation}
 q_{\bar{p}} = \sum\limits_{i,j=p,\text{He}} 4\pi \int dT' \left(\frac{d\sigma}{dT}\right)_{ij} \rho_{j} \Phi_i(T')\,.
\end{equation}
Here we included only processes involving protons and helium, the contribution from heavier nuclei is negligible. The flux and the kinetic energy per nucleon of the incoming primary cosmic ray particle (proton or helium) are denoted by $\Phi_i$ and $T'$, the kinetic energy of the outgoing antiproton by $T$. $\rho_{j}$ stands for the interstellar number density of the target. The differential antiproton production cross section in proton proton scattering $(d\sigma/dT)_{pp}$ is obtained from~\eqref{eq:contributions}. Cross sections involving helium are determined through~\eqref{eq:decomposition} (see~\cite{Kappl:2014hha} for details). We provide the cross sections $(d\sigma/dT)_{pp}$, $(d\sigma/dT)_{p\text{He}}$, $(d\sigma/dT)_{\text{He}p}$, $(d\sigma/dT)_{\text{He}\text{He}}$ in table form for independent use in cosmic ray studies (see ancillary files of the arXiv version).

\begin{table}[htp]
\begin{center}
\begin{tabular}{|ccccc|}
\hline 
\rowcolor{light-gray}&&&&\\[-3mm]
\rowcolor{light-gray} $\delta$ & $K_0\;(\text{kpc}^2\,\text{Myr}^{-1})$ & $L\;(\text{kpc})$ & $V_c\;(\text{km}\,\text{s}^{-1})$& $V_a\;(\text{km}\,\text{s}^{-1})$\\[1mm]
\hline
&&&&\\[-3mm]
$0.408$ & $0.0967$ & $13.7$ & $0.2$ & $31.9$ \\ \hline
\end{tabular}
\end{center}
\caption{Propagation parameters derived in~\cite{Kappl:2015bqa}.}
\label{tab:propagation}
\end{table}

For the propagation of antiprotons, we employ the two-zone diffusion model~\cite{Maurin:2001sj,Donato:2001ms,Maurin:2002ua}. In this scheme diffusion and convection occur homogeneously in a cylinder of half-height $L$. The strength of diffusion is controlled by the parameter $K_0$ and its energy-dependence by the power law index $\delta$. Convection scales with the velocity of the convective wind $V_c$. In addition, energy losses, reacceleration (with the Alfv\'{e}n speed $V_a$), annihilation and inelastic scatterings take place in the galactic disc. The five propagation parameters $K_0$, $\delta$, $L$, $V_c$ and $V_a$ which determine the interstellar antiproton flux $\Phi_{\bar{p}}^\text{IS}$ have to be determined by analysis of nuclear cosmic rays species. We employ the best fit configuration from the boron to carbon analysis~\cite{Kappl:2015bqa} (see table~\ref{tab:propagation}), primary fluxes of protons and helium are taken from the same reference. The measured antiproton flux at the top of the earth-atmosphere $\Phi_{\bar{p}}^\text{TOA}$ is affected by solar modulation for which we account through the force field approximation~\cite{Gleeson:1968zza}, where the value $\phi=0.57\:\text{GV}$ is chosen for the Fisk potential~\cite{Kappl:2015bqa}. Finally we convert kinetic energy into rigidity $\mathcal{R}$ for the comparison with the antiproton flux measured by AMS-02~\cite{Aguilar:2016kjl}.

\begin{figure}[htp]
\begin{center}   
 \includegraphics[width=15cm]{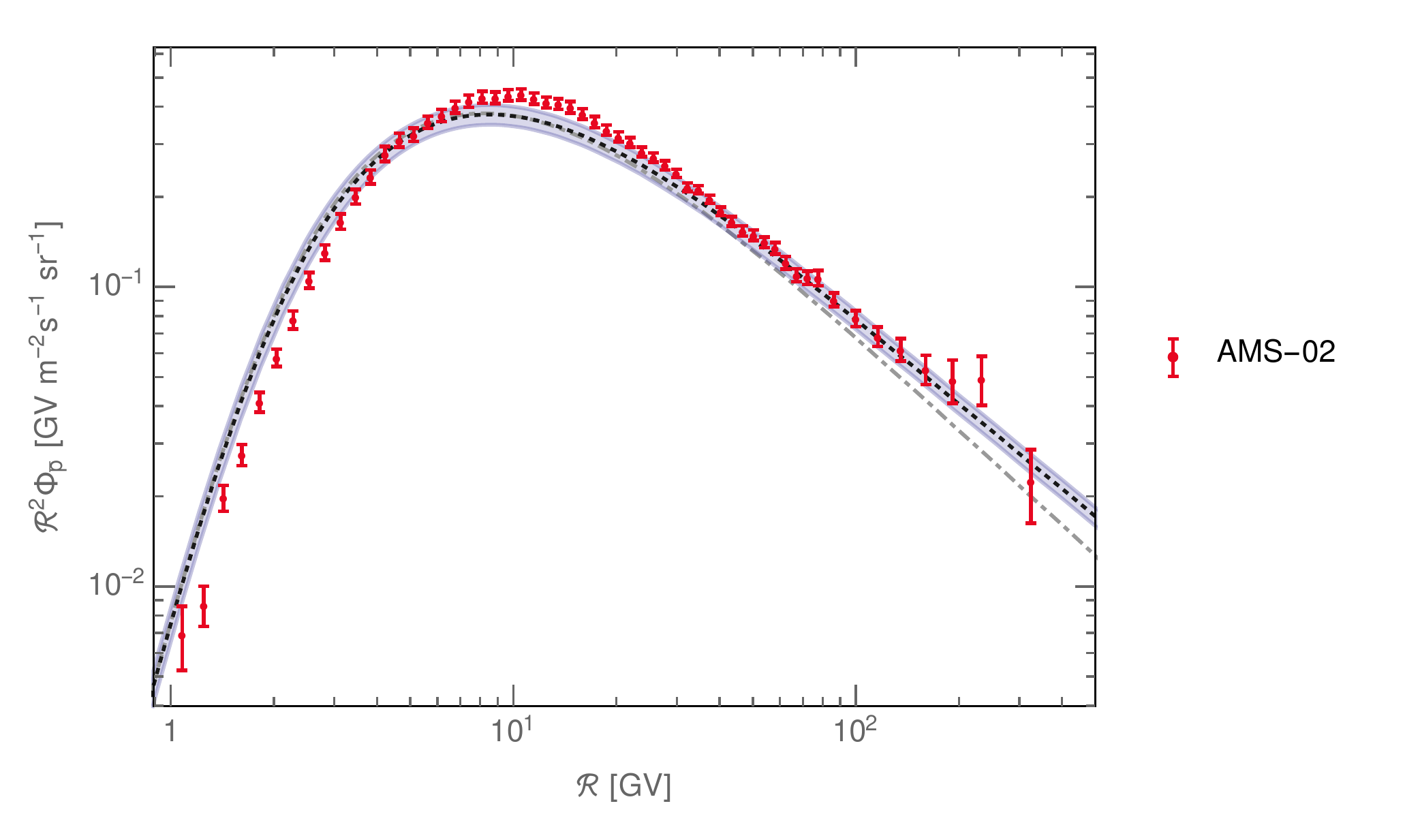}
\end{center}
\caption{Antiproton flux and uncertainty band predicted with the cross sections of this work compared to the flux measured by AMS-02. Also shown is the flux obtained with the previous cross section parameterization~\cite{Kappl:2014hha} (dash-dotted line).}
\label{fig:amsflux}
\end{figure}

In figure~\ref{fig:amsflux} we compare the antiproton flux obtained with our new cross section parameterization to the AMS-02 data~\cite{Aguilar:2016kjl}. We also depict the antiproton flux obtained with our previous parameterization~\cite{Kappl:2014hha}. It can be seen that the new cross sections lead to a harder antiproton flux compared to~\cite{Kappl:2014hha}. The latter results from the effects of radial scaling violation as well as from the increase of hyperon-induced antiprotons with energy. The flatness of the AMS-02 spectrum at high energies is now fully explained and does not require any kind of primary antiproton source in the galaxy. At low rigidity $\mathcal{R}< 5\gev$ our predicted flux exceeds the data, while it lies below the data at intermediate rigidity $\mathcal{R} \simeq 5- 50 \gev$. In order to determine the significance of the deviation, we now turn to a detailed discussion of uncertainties.

\section{Cross Section Uncertainties}

Our cross section parameterization depends on 16 parameters $c_i$ which fix hyperon-induced antiproton production ($c_1-c_4$), prompt antiproton production ($c_5-c_{13}$) as well as the strength of isospin effects ($c_{14}-c_{16}$). We have determined their probability distribution from the available experimental data (the procedure was explained in more detail in section~\ref{sec:hyperon}) and randomly generated tuples \{$c_{1},\dots,c_{16}$\} according to their probability distribution. 
For each of the $N$ cross section configurations (defined by one parameter tuple) we separately calculated the antiproton flux $(\Phi_{\bar{p}})_n$, where the index $n$ runs from 1 to $N$.\footnote{Here and in the following we refer to top-of-the-atmosphere fluxes.} In practice we found sufficiently stable results for $N=500$. The mean predicted flux is defined as $\Phi_{\bar{p}} = \frac{1}{N}\sum\limits_{n=1}^N (\Phi_{\bar{p}})_n$. The half-width of the uncertainty band shown in figure~\ref{fig:amsflux} corresponds to the standard deviation within our sample of antiproton fluxes at a given energy.

AMS-02 has measured the antiproton flux in 57 rigidity bins~\cite{Aguilar:2016kjl} which we number from 1 to 57 with growing rigidity. Each cross section configuration yields a prediction for the flux $(\Phi_{\bar{p},i})_n$ in the $i$-th bin. The mean predicted flux in the $i$-th bin is denoted by $\Phi_{\bar{p},i}$. The cross section uncertainties in the antiproton flux can be given in the form of a covariance
\begin{equation}
 \Sigma_{ij}^\text{cross}= \frac{1}{N}\sum\limits_{n=1}^{N}\Big( \left((\Phi_{\bar{p},i})_n - \Phi_{\bar{p},i}\right) \left( (\Phi_{\bar{p},j})_n - \Phi_{\bar{p},j}\right) \Big)\,,
\end{equation}
where the averaging is performed over the antiproton fluxes within our sample. The diagonal entries of the covariance matrix $\Sigma_{ii}^\text{cross}$ define the standard deviation in the $i$-th bin $\sigma_i^\text{cross} = \sqrt{\Sigma_{ii}^\text{cross}}$. Standard deviations scale with the normalization of the flux and are, therefore, modified if we consider a different set of propagation parameters. Relative standard deviations $\sigma_i^\text{cross}/\Phi_{\bar{p},i}$ are, however, expected to be insensitive to changes in the propagation. We provide $\sigma_i^\text{cross}/\Phi_{\bar{p},i}$ for AMS-02 in the ancillary files of this work (arXiv version). 

A useful quantity closely related to the covariance is the correlation defined as
\begin{equation}\label{eq:correlation}
 \rho_{ij}^\text{cross}=\frac{\Sigma_{ij}^\text{cross}}{\sigma_i^\text{cross}\sigma_j^\text{cross}}\,.
\end{equation}
While the covariance changes with the propagation parameters, the correlation remains (nearly) constant. We, therefore, provide the correlation matrix $\rho^\text{cross}$ for AMS-02 in the ancillary files. For any antiproton flux calculated in future cosmic ray studies, the corresponding covariance matrix of cross section uncertainties can simply be calculated from~\eqref{eq:correlation} by using our published correlation matrix and standard deviations.

Once the covariance matrix of cross section uncertainties is known, the experimental errors of AMS-02 have to be added. The full covariance $\Sigma$ reads
\begin{equation}
 \Sigma_{ij} = \Sigma_{ij}^\text{cross} + \left(\sigma_i^{\text{AMS}} \right)^2\delta_{ij}\,,
\end{equation}
with $\sigma_i^\text{AMS}$ denoting the quadratic sum of statistical and systematic errors in the bin $i$.\footnote{Here we assumed that the experimental errors are uncorrelated.}
We can now perform a $\chi^2$ test of our predicted flux against the AMS-02 data
\begin{equation}
 \chi^2 = \sum\limits_{i,j=1}^{57} \left(\Phi_{\bar{p},i} - \Phi_{\bar{p},i}^\text{AMS}\right)  (\Sigma^{-1})_{ij}  \left(\Phi_{\bar{p},j} - \Phi_{\bar{p},j}^\text{AMS}\right)\,,
\end{equation}
where $\Phi_{\bar{p},i}^\text{AMS}$ denotes the measured flux in the $i$-th bin, while $\Sigma^{-1}$ stands for the inverse of the covariance matrix.

For our predicted antiproton flux we find $\Delta\chi^2/\text{d.o.f.} = 1.4$ (ignoring cross section uncertainties would have led to $\Delta\chi^2/\text{d.o.f.}=8.5$). The fit is considerably better than figure~\ref{fig:amsflux} suggests. The reason is that cross section uncertainties are highly correlated which could not be made visible in the figure. While the value of $\Delta\chi^2$ formally corresponds to an exclusion at $2.1\,\sigma$, we point out that we did not consider all relevant sources of error. Including uncertainties in the propagation and in the solar modulation of antiprotons is expected to further improve the quality of the fit.

\section{Conclusion}

In this article we have reevaluated the antiproton production cross sections which enter the calculation of the cosmic ray antiproton flux. Compared to previous studies we have systematically analyzed proton proton scattering data of RHIC and LHC to identify the effects of scaling violation in the cross sections. Besides in the growth of the inelastic cross section, they manifest themselves in the transverse momentum distribution of antiprotons. The transition from an exponential to a power law dependence on the transverse antiproton mass is described by Tsallis statistics. Scaling violations are found to increase the antiproton cross sections at high energies and lead to a harder antiproton spectrum. In addition, experimental data on $\bar{\Lambda}$ production prove an increase of the strange hyperon multiplicity relative to the antiproton multiplicity towards high energies. This, in term, triggers a further hardening of the cosmic ray antiproton flux through the decay of energetic hyperons. Finally, we employed proton proton, neutron proton and proton nucleus scattering data to determine the strength of isospin effects which induce an asymmetry between antiproton and antineutron production. While the presence of isospin effects is favored by low energy data, they disappear towards high energies. We determined the energy-dependence of isospin effects and their impact on the antiproton production. Our newly determined cross sections for proton proton, proton helium, helium proton and helium helium scattering are made publicly available in the form of tables.

The secondary cosmic ray antiproton flux calculated with the new cross sections was then compared to the AMS-02 data. Thereby, we kept track of all relevant sources of cross section uncertainties and determined how they propagate into uncertainties of the flux. With this article we publish the cross section uncertainties in each rigidity bin of AMS-02 as well as their full correlation matrix. These can independently be used in future likelihood analyses of the AMS-02 antiproton data. In particular, the cross section uncertainties of the secondary background are a crucial input for dark matter searches using antiprotons.

Performing a likelihood test over the full energy range we found that the AMS-02 antiproton data are consistent with a pure secondary origin at the level of $\sim 2 \sigma$. The fit is expect to further improve, once uncertainties in the propagation of cosmic rays as well as in the solar modulation are included. The rise of the antiproton cross sections provides the final piece in understanding the hard antiproton spectrum observed by AMS-02 at high energies.

\section*{Acknowledgments}
I would like to thank Rolf Kappl and Annika Reinert for helpful discussions.

\bibliography{antibib}
\bibliographystyle{ArXiv}

\end{document}